\documentclass[aps,pre,showpacs,twocolumn,superscriptaddress,
              floats,floatfix]{revtex4-1}
\usepackage{amssymb,amsmath}
\usepackage{graphics}
\usepackage{color}
\usepackage{subfigure}
\usepackage{epsfig}
\usepackage{bm}
\usepackage{ulem}
\graphicspath{{./fig/}{./png/}}

\def \uu {{\bm u}}
\def \Uzero {{U_0}}
\def \Ustar {{U_{\ast}}}
\def \Uplus {{U^{+}}}
\def \ystar {{y_{\ast}}}
\def \yplus {{y^{+}}}

\newcommand{\eq}[1]{(\ref{#1})}

\newcommand{\Eq}[1]{Equation~(\ref{#1})}
\newcommand{\Eqs}[2]{Equations~(\ref{#1}) and~(\ref{#2})}

\newcommand{\eqs}[2]{(\ref{#1}) and~(\ref{#2})}
\def \RR  {{\bm R}}

\def \hate {{\bm e}}
\def \dxj {\delta {\bm x}^{\rm j}}
\def \dxx {\delta {\bm x}}
\def \np {N_{\rm p}}
\def \Rmax {R_{\rm max}}
\def \rjt {{\bm r}^{\rm j}(t|t_0,{\bm r}^j_0)}

\def \vjt {{\bm v}^{\rm j}(t|t_0,{\bm r}^j_0)}

\def \grad {{\bm \nabla}}

\def \dive {{\bm \nabla}\cdot}
\def \lap {\nabla^2}
\def \delt {\partial_t}

\newcommand{\bra}[1]{\left\langle #1\right\rangle}
\def \Rey  {\mbox{Re}}
\def \Wi  {\mbox{Wi}}

 \def \taup {\tau_{\rm poly}}
 \def \tauf {\tau_{\rm fluid}}
 \def \tauL {\tau_{\rm L}}
 \def \mubar {\bar{\mu}}
 \def \muT {\mu_{\rm T}}
 \def \yplus {y^{+}}
 \def \uplus {U^{+}}

\begin{document}
\title{Statistics of polymer extensions in turbulent channel flow} 
\author{Faranggis Bagheri}
\email{fbagheri@kth.se}
\affiliation{Linn\'e Flow Centre, KTH Mechanics, SE-100 44 Stockholm, 
       Sweden}
\author{Dhrubaditya Mitra}
\email{dhruba.mitra@gmail.com}
\affiliation{NORDITA, Roslagstullsbacken 23, 106 91 Stockholm,
Sweden}
\author{Prasad Perlekar}
\email{p.perlekar@tue.nl}
\affiliation{Department of Mathematics and Computer Science, 
Eindhoven University of Technology, P.O. Box 513, 5600 MB Eindhoven, 
The Netherlands}
\author{Luca Brandt}
\email{luca@mech.kth.se}
\affiliation{Linn\'e Flow Centre, KTH Mechanics, SE-100 44 Stockholm, 
       Sweden}
\begin{abstract}

We present direct numerical simulations of turbulent channel flow 
with passive Lagrangian polymers.   
To understand the polymer behavior we investigate the behavior of 
infinitesimal line elements and calculate the probability distribution
function (PDF) of finite-time Lyapunov exponents and from them the corresponding 
Cramer's function for the channel flow.
We study the statistics of polymer elongation for both the Oldroyd-B 
model (for Weissenberg number $\Wi <1$ ) and the FENE model. 
We use the location of the minima of the Cramer's function to define the Weissenberg 
number precisely such that we observe coil-stretch transition
at $\Wi\approx1$. 
We find agreement with earlier analytical predictions for PDF of 
polymer extensions made by Balkovsky, Fouxon and Lebedev
[Phys. Rev. Lett. {\bf 84}, 4765 (2000).] for  linear polymers (Oldroyd-B model)
with $\Wi<1$  and by Chertkov [Phys. Rev. Lett. {\bf 84}, 4761 (2000).] for nonlinear FENE-P model of polymers. 
For $\Wi>1$ (FENE model) the polymer are significantly more stretched 
near the wall than at the center  of the channel where the flow is closer to  
homogenous isotropic turbulence.
Furthermore
near the wall the polymers show a strong tendency to orient
along the stream-wise direction of the flow but near the centerline
the statistics of orientation of the polymers is consistent with
analogous results obtained recently in homogeneous and isotropic 
flows.
\end{abstract}
\maketitle 
\section{Introduction}
Turbulent flows with polymer additives have been an active field of
interest since the discovery~\cite{toms49}  of the  
phenomenon of drag reduction on the addition of small 
amounts (few parts per million) of long-chained polymers to turbulent 
wall-bounded flows. 
Polymers are long-chained complex molecules which have roughly
spherical equilibrium configurations, known as the ``coiled'' state.
In the simplest models of polymers, the relaxation of the polymers
to the coiled state can be described by a single time scale  
$\taup$.
If such a polymer is then placed in a straining flow, where
the strain can be characterised by inverse of a time scale
$\tauf$,  the polymer can go from its coiled state
to a stretched state if the ratio of the two time scales,
the Weissenberg number $\Wi > 1$~\cite{squ05}. 
Thus in turbulent flows with strong strain the polymers can go through 
coil-stretch transition; the stretched polymers can then make significant 
contribution to the Reynolds stresses and this can result in
drag-reduction~\cite{lum73,hin77}. 
Hence to understand drag-reduction
we must first understand the mechanism of coil-stretch transition.
Also note that the back-reaction of the polymers to the flow becomes
significant only when the polymers have undergone coil-stretch transition,
thus to study coil-stretch transition itself, it may be safe to consider passive 
polymers. 

There has been a large volume of work on coil-stretch transition of polymers
in various kinds of flows. These works can be divided in four classes depending on the properties of the
flow:
(A) Individual polymer molecules advected by synthetic flows.
In this class we first mention analytical works where the flows are
either assumed to be random, smooth and white-in-time, Batchelor-Kraichnan 
flows~\cite[see e.g.,][]{deu92,bal+fux+leb00,che00,thi03,afo+vin05,tur07}, or
to have simple prescribed time dependence~\cite[e.g., Refs.][]{lum73,mas+kon+sch+han93,mus+vin11}.  For $\Wi < 1$ the
analytical works in  Batchelor--Kraichnan flows have predicted that the probability
distribution function (PDF) of polymer extension exhibits a power-law tail. 
Next are numerical works where the PDF of polymer extension and polymer tumbling times 
are calculated for polymers in various synthetic flows, including Batchelor-Kraichnan flows superimposed on 
uniform shearing background~\cite{cel+pul+tur05,che+kol+leb+tur05} and models of a turbulent buffer layer ~\cite[e.g.,][]{sto+gra03}. 
(B) Lagrangian polymers advected by solutions of the Navier--Stokes
equation\cite[see e.g.,][]{mas+kon+sch+han93,eck+kro+sch02,gup+sur+kho04,wat+got10}. 
(C) Numerical simulations where the equations of polymers and fluids
are solved simultaniously in two~\cite{bof+cel+mus03,gup+per+pan12} and three~\cite[see e.g.,][]{per+mit+pan06,per+mit+pan10,dal+vas+hew10} spatial dimensions.
(D) And finally numerical simulations of Lagrangian polymers in solutions of Navier--Stokes equations, 
in which the back-reaction from the polymers to the fluid is attempted to be incorporated~\cite{dub04,Ter05,pet07}.

The simplest analytically tractable model is that of class (A) above.  In this model the polymer is described by
a simple bead-spring model, 
\begin{equation}
\delt R_{\alpha}(t) = \sigma_{\alpha\beta}R_{\beta} + f(R) 
\label{eq:polyone}
\end{equation}
where $\RR$ is the end-to-end vector of a polymer (macro)molecule, $\sigma_{\alpha\beta}$ is a model for
the velocity gradient matrix of the flow, $f(R)$ is the restoring force of the (entropic) spring in the
bead-spring model,  e.g.,$f(R) = - R/\taup$ for a harmonic overdamped spring (Oldroyd-B model), and $\taup$ the characteristic
relaxation time of the polymer. The phenomenon of polymer stretching in flows is best understood by, for a moment, ignoring the restoring
force in \eq{eq:polyone}.  The resulting equation is then the same
equation as the one that describes the evolution of the 
infinitismal separation ($\delta {\bm x}$) between two fluid particles,  i.e.,
\begin{equation}
\partial_t \delta x_{\alpha} = 
         \sigma_{\alpha\beta} \delta x_{\beta}.
\end{equation}
How two infinitesimally separated Lagrangian particles diverge in a
turbulent flow has been a central topic in turbulence research
for a long time. See, e.g., Ref~\cite{fal+gaw+var01} for a recent review. 
Below we reproduce the essential points needed
to apply such ideas to stretching of polymers in turbulence. 

The growth (or decay) of the distance between two Lagrangian particles up to a time $T$ is
described by the finite-time-Lyapunov-exponents(FTLEs) defined by 
\begin{equation}
\muT = \frac{1}{T}\ln \left[ \frac{|\dxx(t)|}
                      {|\dxx(t-T)|}  \right].
\end{equation}
For large $T$, $T\to \infty$ the PDF of FTLEs is conjectured to have a large deviation
form~\cite{bal+fux+leb00,eck+has+bra03,gaw08}
\begin{equation}
P(\mu_T) \sim \exp[-T S(\mu_T) ]
\label{eq:cramer}
\end{equation}
where $S(\mu_T)$ is called the Cramer's function or the entropy function.
The simplest form of the entropy function is a parabola, of the form 
 $S(\mu) = (\mu-\mubar)^2/\Delta$,
in which case (for each time $T$)
the PDF of the FTLEs is a Gaussian distribution. The mean value of this Gaussian ($\mubar$) is an inverse
time scale, $\mubar = 1/\tauf$.  For a turbulent (or random) flow the Weissenberg number is best defined
by the ratio of $\taup/\tauf$.  The analytical work of Ref.~\cite{bal+fux+leb00} calculated the
the PDF of polymer extensions in a random homogeneous flows with short correlation time.  
They found that for $\Wi < 1$ the PDF has power-law tail with an exponent $\alpha$.
This exponent $\alpha$ can be obtained from the Cramer's function
$S(\mu)$ by solving the following set of coupled equations Eq.~(\ref{eq:alpha}) and 
(\ref{eq:beta}) given below,
\begin{eqnarray}
\alpha &=& S^{\prime}(\beta+\frac{1}{\taup}-\mubar) 
\label{eq:alpha} \\
S(\beta+\frac{1}{\taup}-\mubar) &-& \beta S^{\prime}(\beta+\frac{1}{\taup}-\mubar) = 0
\label{eq:beta}
\end{eqnarray}
Had the Cramer's function been well approximated by a parabola of the
form $S(\mu) = (\mu-\mubar)^2/\Delta$, Eq.~(\ref{eq:alpha}) would
simplify to $\alpha = (2/\Delta)(1/\taup-\mubar)$. 
 Let us state here explicitly the assumptions that goes behind the derivation of 
\Eqs{eq:alpha}{eq:beta}. 
The velocity gradient matrix is assumed to be short correlated in time, smooth in space
and invariant under three dimensional rotation. In addition it is
assumed that the PDF of FTLEs 
having a large deviation form holds true.  Our numerical work, presented below, 
shows that both of these assumptions can be made for a three dimensional channel flow. 

In this paper we calculate, the PDF of finite-time-Lyapunov exponents, for both short and large times, 
in turbulent channel flow by direct numerical simulation (DNS).
 We then show that at large time the PDF of FTLEs does satisfy a
large deviation form with a Cramer's function that can be approximated by a fourth--order polynomial. 
We further solve for the Oldroyd-B model of Lagrangian polymers in this flow. 
We use the location of the minima of the Cramer's function, $\mubar$,
as the inverse characteristic time scale of the fluid to define our 
Weissenberg number as 
\begin{equation}
\Wi \equiv \mubar\taup
\label{eq:wi}
\end{equation}
Our simulations show a coil-stretch transition for $\Wi \simeq 1$. 
For $\Wi<1$, the PDF of polymer
extension shows a power-law tail with scaling exponent $\alpha$. 
We find that the range of scaling  shown by the PDF of polymer extensions depends on 
the wall-normal  coordinate but the scaling exponent $\alpha$ is independent of the 
wall-normal coordinate. We further show that the exponent $\alpha$ satisfies
\eqs{eq:alpha}{eq:beta}.

For $\Wi > 1$ it is not possible to obtain a stationary PDF for
the Oldroyd-B model. In this regime
we use the nonlinear FENE (Finitely Extendable Nonlinear 
Elastic) model.  For this model analytical work~\cite{che00}
has found that
\begin{equation}
\bra{\frac{R}{\Rmax}} = -\frac{1}{\mubar}f(\bra{R}) .
\end{equation} 
Our numerical simulations confirm this result. 
In addition we also find that the PDF of polymer extensions depends  
on the wall-normal coordinate, v.i.z, the polymer are more stretched 
near the wall than at the center of the flow. We further study
the orientation of the polymers with respect to the channel 
geometry and the local velocity gradient tensor.  Our results show 
that the orientation of the polymers is predominantly determined by 
the inhomogeneity of the flow, i.e., by the wall-normal coordinate as 
opposed to the local strain tensor. 
However, for polymers near the center of the channel
we find that the orientation is also influenced by the principal
directions of the rate-of-strain tensor, as has been seen
in DNS of polymers in homogeneous isotropic 
flows~\cite{jin+col07,wat+got10}.

The rest of the paper is organised as follows.  In  Section~\ref{sec:model} we describe the equations we
solve and the details of the numerical algorithm we use.  
Our results follow in
Section~\ref{sec:results} which is divided into three parts. 
The PDF of the finite-time-Lyapunov-exponents (FTLEs) are reported in Section~\ref{subsec:FTLE}.
The results described in 
this section are therefore independent of the polymer equation. 
The statistics of polymer extensions for the 
two models considered are presented in Section~\ref{subsec:PDF} 
and~\ref{sub:FENE}. The polymer orientation is characterized by calculating 
the correlations between the polymer end-to-end vector 
and fluid vorticity and the rate of strain 
tensor (Section~\ref{subsec:angle}). 
The main conclusions of the study are summarized in Section~\ref{sec:conclusion}.

\section{Equations and Numerical Methods} 
\label{sec:model}
The fluid is described by the Navier--Stokes
equations, 
\begin{equation}
\partial_t \uu + \uu\cdot\grad \uu = \nu \lap \uu + \grad p 
\label{eq:NS}
\end{equation}
with the incompressibility constraint,
\begin{equation}
\dive \uu = 0.
\label{eq:incompress} 
\end{equation}
Here $\uu$ is the fluid velocity, $\nu$ the kinematic viscosity, and $p$ the
pressure.
We use the no-slip boundary condition at the walls and periodic boundary 
condition at all other boundaries. 
We have chosen our units such that the constant density
$\rho = 1$. The $x$ axis of our coordinate system is taken along
the stream-wise direction, the $y$ axis along the wall-normal
direction and $z$ axis along the span-wise direction.
For brevity, we shall often use the common notation, $U \equiv u_x$, 
$V\equiv u_y$ and $W\equiv u_z$. 
The $x$, $y$ and $z$ dimensions of our channel are
$L_x\times L_y \times L_z = 4\pi\times 2\pi \times 2\pi$,
with resolution 
${\tt Nx}\times{\tt Ny}\times{\tt Nz} = 128\times 129 \times 128$. 

The turbulent Reynolds number $Re_{\tau} = \Ustar L/\nu =180$ is defined by 
the friction velocity $\Ustar = \sqrt{\sigma_{\rm w}}$ and $L\equiv L_y/2$, 
the half-channel width, where 
\begin{equation}
\sigma_{\rm w} \equiv \nu \frac{\partial U}{\partial y} |_{\rm wall}
\end{equation} 
is the shear stress at the wall~\cite{Mon+Yagv1}. 
In the following we non-dimensionalise velocity and distance by 
$\Uplus\equiv U/\Ustar$ and $\yplus\equiv y/\ystar$ respectively, using
the friction length $\ystar=\nu/\Ustar$. 
Time will also be measured in the unit of 
the large-eddy-turnover-time 
$\tauL \equiv (U^{\rm center}/L)^{-1}$, 
where $U^{\rm center}$ is average streamwise velocity at the center of the channel.
The large-scale Reynolds number defined by 
$\Rey = \Uzero L/\nu= 4200$ 
where $\Uzero$ is the centreline streamwise velocity for the laminar 
flow of same mass flux.

\begin{figure}[!h]
\begin{center}
\includegraphics[width=0.9\columnwidth]{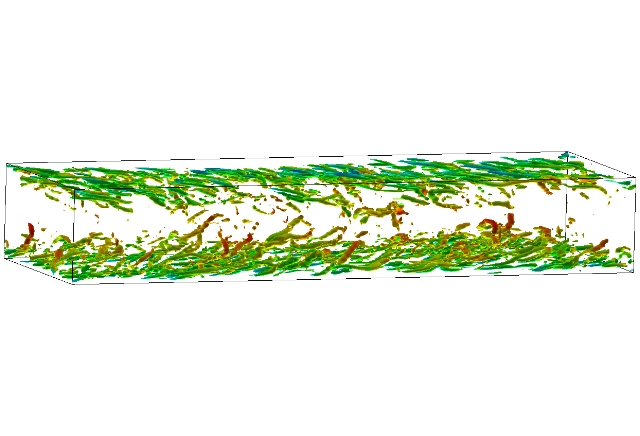}
\put(-77,40) {\Large{(a)}} \\
\includegraphics[width=0.9\columnwidth]{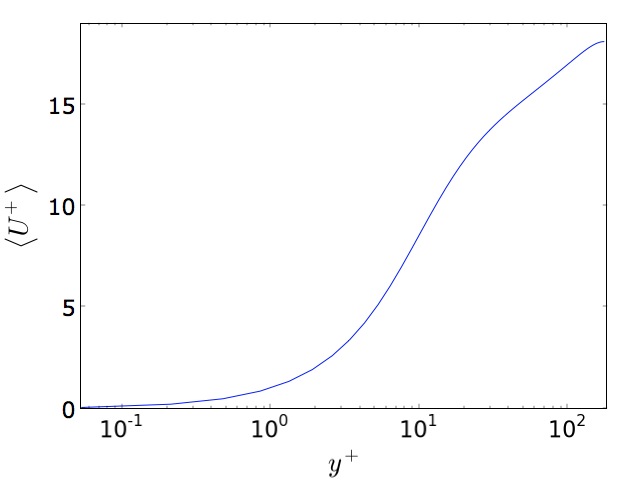}
\put(-77,40) {\Large{(b)}} \\
\caption{\label{fig:ux}(Colour online)
(a) Visualization of vortical structures in our simulation. 
The structures are identified by negative values of the second largest
eigenvalue of the matrix $S_{ik}S_{kj} + \Omega_{ik}\Omega_{kj}$
where $S_{ij}$ and $\Omega_{ij}$ and the symmetric an anti-symmetric part
of the velocity gradient matrix~\cite{jeo+hus+sch+kim97}.  
The vortical structures are located close to the walls.
(b) Normalised mean stream-wise
velocity $\bra{\uplus} $ versus the wall-normal coordinate $\yplus$ plotted
for one half of the channel.} 
\end{center}
\end{figure}

We solve Eqs. (\ref{eq:NS}) and (\ref{eq:incompress}) by using
the SIMSON~\cite{SIMSON} code which uses a pseudo-spectral method 
in space (Chebychev-Fourier). 
For time integration a third-order Runge-Kutta method is used
for the advection term and the uniform pressure gradient term. The viscous 
term is discretized using a Crank-Nicolson method~\cite{mos+kim+man99}. 
In Fig.~(\ref{fig:ux}a) we show a visualization of
the vortical structures from a typical snapshot of our simulation. 
In Fig.~(\ref{fig:ux}b) we plot $\bra{\Uplus}$ 
as a function of $\yplus$ at the stationary state of our 
simulations, where $\bra{\cdot}$ denotes averaging over the 
coordinate directions $x$ and $z$ and over time.  Further details about the 
code validation can be found in Ref.~\cite{Far10,sar+sch+bra+pic+cas12}. 

We use a Lagrangian model for the polymers where we solve one stochastic
differential equation (SDE) for each polymer molecule. This model uses several
approximations which are as follows~\cite{Bir87,Pha02}: (A) The
centre-of-mass of a polymer molecule follows the path of a Lagrangian
particle. (B) Even when fully stretched the polymer molecule
is very small compared to the smallest scales of turbulence. This
approximation is well justified~\cite{hin77}. 
(C) A polymer molecule is modelled by two beads separated by a
vector which represents the end-to-end distance of the polymer
molecule. (D) The forces acting on the beads are stokes drag, restoring force
of an overdamped spring with time scale $\taup$, and thermal noise. 
To be specific, we track $\np=2.16\times10^5$ Lagrangian passive tracers
in the flow by solving 
\begin{equation}
\delt \rjt = \vjt \/.
\label{eq:rlag}
\end{equation}
Where $\rjt$ is the position of the $j$-th Lagrangian particle 
which was at position ${\bm r}^{\rm j}_0$ at time $t_0$ and
$\vjt$ is its velocity with $j=1,\ldots,\np$. 
The Lagrangian velocity
of a particle, which is generally at an off-grid point, is obtained by 
tri-linear interpolation from Eulerian velocity at the neighboring 
grid points.
Equation~(\ref{eq:rlag}) is integrated by a third order Runge-Kutta scheme.
Each of these Lagrangian particles represent a polymer molecule. 
For $j$-th Lagrangian particle the vector representing
the end-to-end distance is denoted by $R^{\rm j}$ and obeys the following
dynamical equation:
\begin{equation}
\delt R^{\rm j}_{\alpha}(t) = \sigma^{\rm j}_{\alpha\beta}R^{\rm j}_{\beta} + 
                  f(R^{\rm j}) + 
                   \sqrt{\frac{2R^2_0}{3\taup}}B^{\rm j}_{\alpha}\/.
\label{eq:poly}
\end{equation}
Here $\sigma^{\rm j}_{\alpha\beta} = \partial_{\beta}\vjt_{\alpha}$, $f(R^{\rm j})$ is 
the restoring force of the polymer, $\taup$ is the characteristic decay
time of the polymer  and ${\bm B}^{\rm j}$ is a Gaussian random noise with
$\bra{B_{\alpha}}=0$ and 
$\bra{B_{\alpha}(t)B_{\beta}(t^{\prime})}=\delta_{\alpha\beta}\delta(t-t^{\prime})$.
The prefactor of the random noise is chosen such that in the absence of 
external flow, 
i.e., $\sigma^{\rm j}_{\alpha\beta} = 0$, the polymer attains thermal
equilibrium,  
$\bra{R^{\rm j}_{\alpha}R^{\rm j}_{\beta}}
     =R^2_0\delta_{\alpha\beta}/3$.
Here $\bra{\cdot}$ denotes averaging over the noise ${\bm B}$. 
For the linear Oldroyd-B model $f(R)=-R/\taup$. 
For the FENE model $f(R)=-R/\taup\{1-(R/\Rmax)^2\}$. Eq.~(\ref{eq:poly})
is also solved by a third order Runge-Kutta scheme except for the noise
which is integrated by an Euler-Maruyama method~\cite{hig01}.

To compare with the analytical theory of Ref.~\cite{bal+fux+leb00} we also need
to calculate the PDF of finite-time Lyapunov exponents of Lagrangian
particles in this flow. For this we need to calculate the rate at which 
two infinitismally separated Lagrangian particles diverge as time
progresses. For this purpose we also calculate the evolution of an infinitesimal vector 
in our turbulent flow, given by the equations,
\begin{equation}
\partial_t \delta x^{\rm j}_{\alpha} = 
         \sigma^{\rm j}_{\alpha,\beta} \delta x^{\rm j}_{\beta}.
\label{eq:deltax}
\end{equation}
Where $\dxj$ is a vector carried by the $j$-th Lagrangian particle.
This is of course the same equation obeyed by a Lagrangian polymer,
[Eq.~(\ref{eq:poly})], if the restoring force of the polymer and the
Brownian noise are omitted.

The correspondence between  
our Lagrangian description and the Eulerian description of 
polymeric fluids is that in the latter the dynamical 
variable for the polymers is the 
symmetric positive definite (SPD) tensor
$C_{\alpha\beta} \equiv \bra{R_{\alpha}R_{\beta}}$.  
A DNS of the Eulerian description has certain 
difficulties~\cite{vai03,vai06,per+mit+pan06,per+mit+pan10}.
Firstly the numerical schemes used must preserve the 
symmetric and positive-definite (SPD) nature of $C_{\alpha\beta}$. 
Secondly for high Weissenberg numbers large gradients of $C_{\alpha\beta}$
can develop which can lead to numerical instabilities. 
Stability can generally be restored by employing either shock-capturing 
schemes~\cite{vai06,Per09,per+mit+pan10,dal+vas+hew10} or by introducing dissipation in 
the Eulerian description of the polymer~\cite{ben03,ang05}.
Lagrangian methods~\cite{dub04,Ter05} are generally able to avoid such numerical 
pitfalls and can attain a higher Weissenberg number. 
On the other hand it is quite straightforward to incorporate the back-reaction 
of the polymer into the flow in the Eulerian model but is tricky 
in the Lagrangian model~\cite{Ter05,pet07}. 
Note finally that more complicated Lagrangian models have also been employed
where a single polymer is represented by a chain of beads connected
by springs~\cite{jin+col07,wat+got10}.

\section{Results}
\label{sec:results}
 
\subsection{Finite-time Lyapunov Exponents}
\label{subsec:FTLE}
To calculate the PDF of FTLEs we integrate Eq.~(\ref{eq:deltax}) for each Lagrangian 
particle over a finite time interval $T$  and compute the
the finite-time Lyapunov exponent~(FTLE) as 
\begin{equation}
\muT^{\rm j}  = \frac{1}{T}\ln \left[ \frac{|\dxj(t)|}
                      {|\dxj(t-T)|}  \right].
\end{equation}
Below we present the analysis
of the PDF of FTLEs for channel flows. 

\begin{figure}
\begin{center}
\includegraphics[width=0.9\columnwidth]{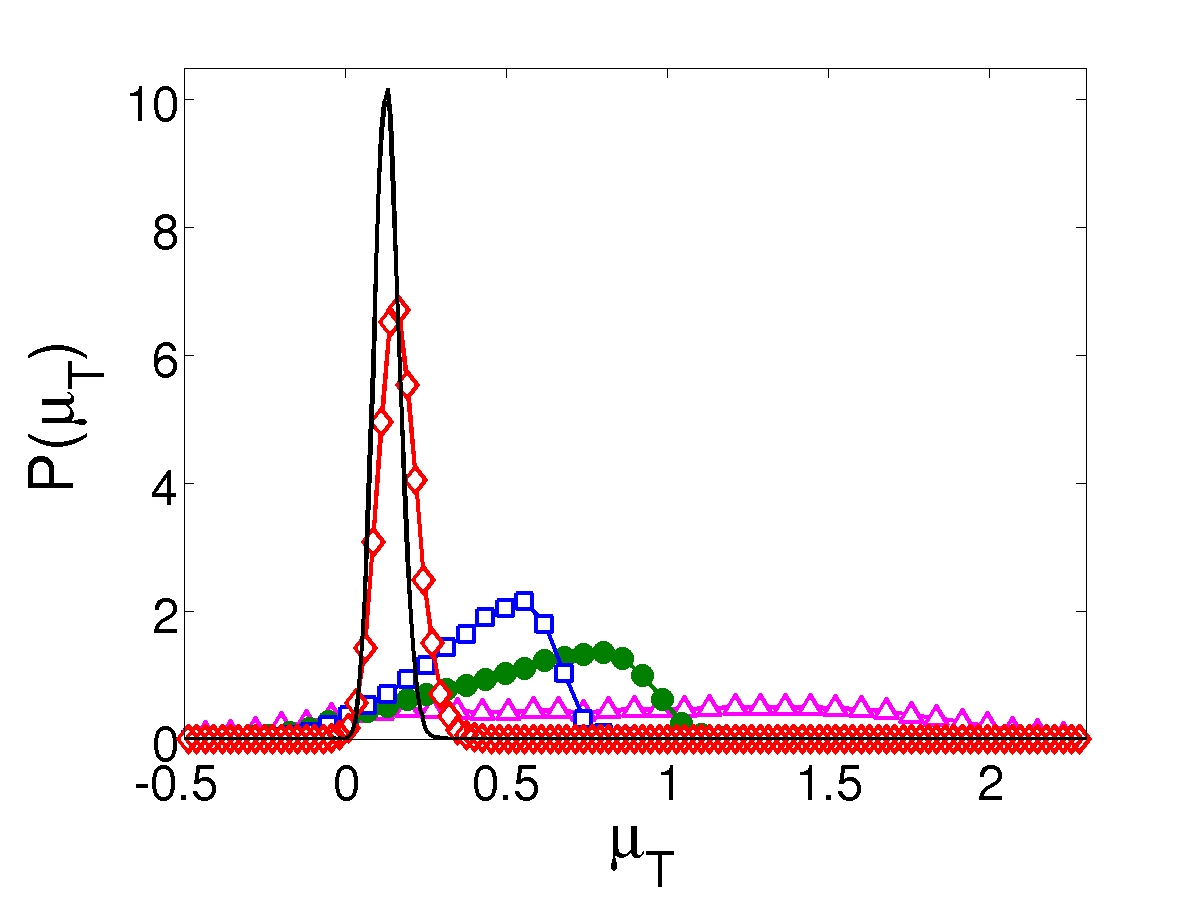}
\put(-50,110) {\Large{(a)}} \\
\includegraphics[width=0.9\columnwidth]{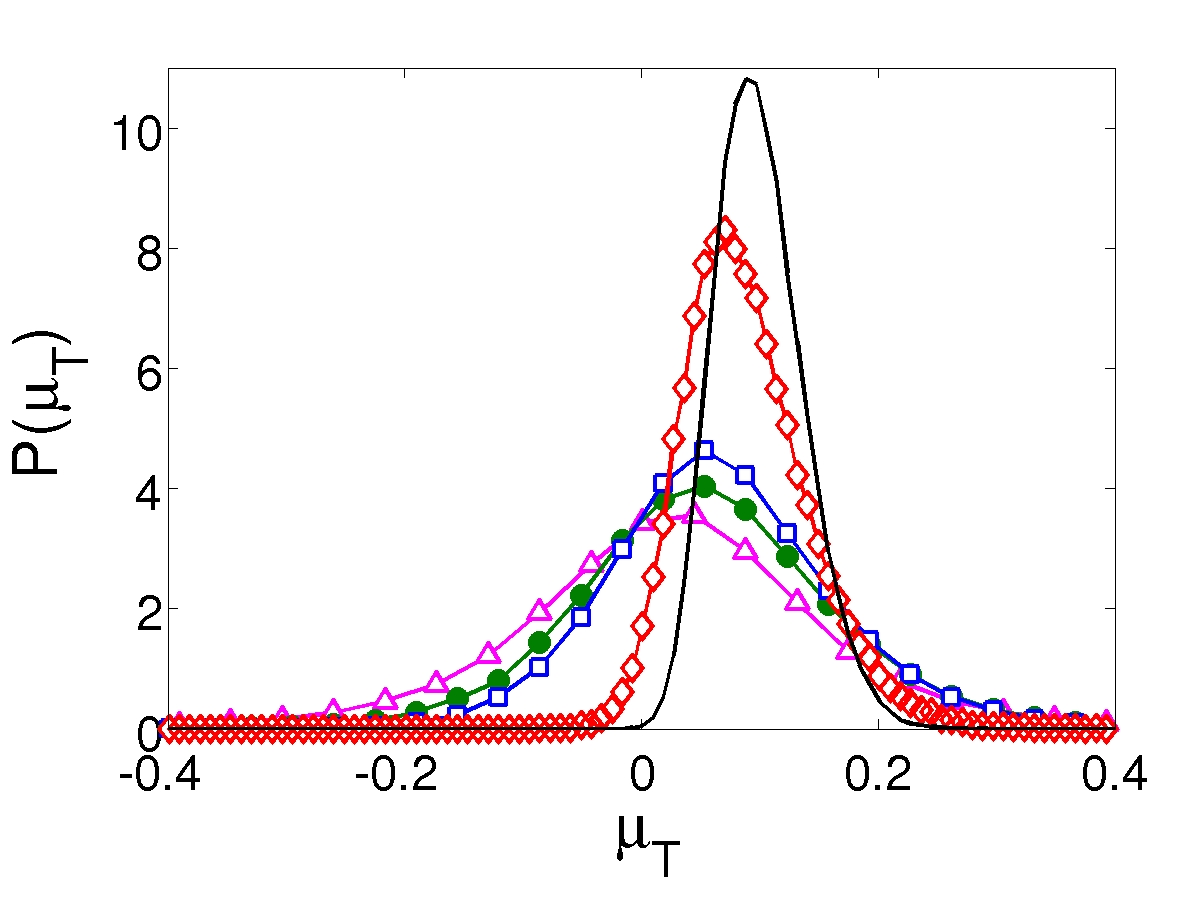}
\put(-50,110) {\Large{(b)}} \\
\caption{\label{fig:pmuwall}(Colour online) 
(a) PDF of $\muT$ near the wall
($\yplus \approx 6$) for several values of $T$, v.i.z, 
$T= 1.(\vartriangle)$, $3.(\bullet)$, $5.(\square)$,
$35. (\lozenge)$,  and $100. (-)$. 
(b) PDF of $\muT$ near the
centerline ($\yplus \approx 180$) for several values of $T$, v.i.z, 
$T= 1.(\vartriangle)$, $3.(\bullet)$, $5.(\square)$, 
$35. (\lozenge)$, and $100. (-)$.
Plots at other intermediate values of $T$ are consistent with
this plot, but are not show here for clarity.  
All times are measured in the unit of $\tauL$.
}
\end{center}
\end{figure}

\begin{figure}[!h]
\begin{center}
\includegraphics[width=0.9\columnwidth]{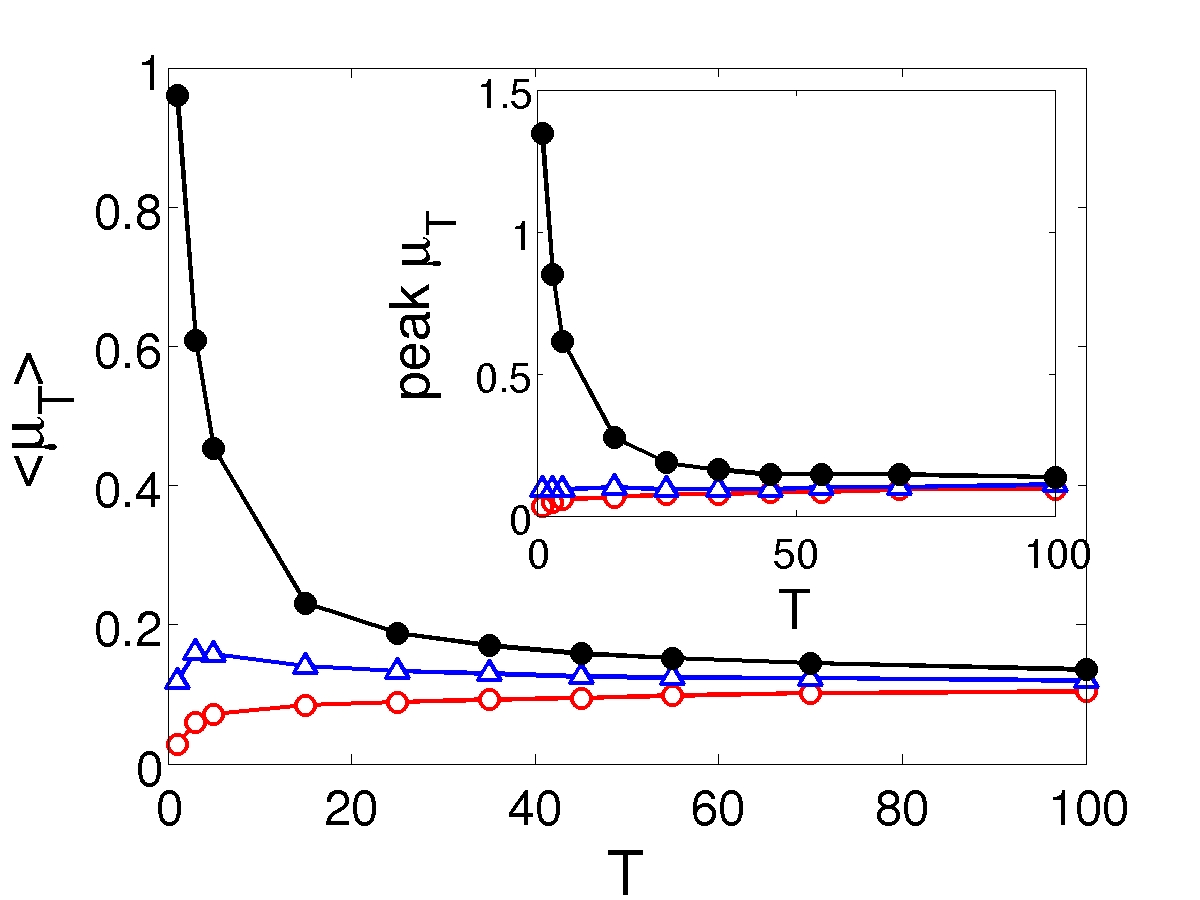}
\caption{\label{fig:meanpeak}(Color online) Mean FTLE $\bra{\muT}$ versus $T$ for three different 
positions, in the channel, near the wall(\textbullet), 
near centerline ($\circ$) and at $\yplus=84$ ($\vartriangle$). 
All times are measured in the units of $\tauL$. }
\end{center}
\end{figure}

Since the channel flow is not homogeneous in the wall-normal direction the 
statistics can, in principle, depend on $y^+$.
Hence we label our particles by 
their wall-normal coordinate ($\yplus$) at the final position, i.e., at time $T$. 
While integrating the equations for $\dxj$, Eq.~(\ref{eq:deltax}), 
we store the evolution of $\dxj$ and use this to 
calculate $\muT^{\rm j}$ for each of $\dxj$. 
To calculate the PDF of $\muT$ we gather statistics in two different ways. 
First we calculate the PDF of $\muT$ for all particles at a fixed $\yplus$. 
Furthermore we run our simulations over several $T$ and after each time 
interval $T$ the particles are redistributed uniformly across the channel
and their initial separation vector $\dxj(t=0)$ oriented randomly. 
By definition then we generate a $P(\muT,\yplus)$ which depends
on $\yplus$. 
The PDFs for two different values of $\yplus$, one close to the
wall, and one near the centerline, are respectively plotted in 
Figs.~(\ref{fig:pmuwall}a)~and~(\ref{fig:pmuwall}b)    
for several time intervals $T$.
The peak and mean of the PDFs are always positive showing
that it is more probable for $|\dxj|$ to increase exponentially
as a function of time.
For small $T$ the PDFs near the center and the PDF near the wall
are very different from each other. Significantly larger 
elongation is found for those elements that are located closer to the wall. 
However, the two PDFs approach each other  for large $T$.  This can also be seen by 
plotting the mean value of the PDFs for three different $\yplus$s as a function of 
time, Fig.~(\ref{fig:meanpeak}). The peak value also shows a similar trend, see the 
inset in Fig.~(\ref{fig:meanpeak}).
Hence an unique Cramer's function independent of $\yplus$ can be defined
for the channel flow for only very large time when the PDFs for different
$\yplus$ merge with one another.  
In a channel flow the stress tensor $\sigma_{\alpha\beta}$ depends strongly on 
the wall-normal coordinate.
Thus for short $T$ we can expect that the PDF of $\muT$ depend of $\yplus$.
Conversely, when $T$ becomes much larger than the typical time it takes for 
a particle to travel from a position near the wall to a position near the center line, 
we expect $\muT$ to be independent of $\yplus$. 
Let us call this typical time the exit time $T_{\rm exit}$. 
Surprisingly, we observe from our data that we need to have 
$T_{\rm exit} \gtrsim 80 \tauL$ for $\muT$ to be independent of $T$. 
An estimate of the time it takes for a particle to travel from the
wall to the center of the channel can be given by the ratio
of the half-width of the channel to the friction velocity,  
$T_{\rm friction} \equiv (L_y/2)/\Ustar \approx 15 $ in our simulations. 
In units of this time $T_{\rm exit} \gtrsim 5 T_{\rm friction}$
which provides a better estimate than $\tauL$. 

\begin{figure}[!h]
\begin{center}
\includegraphics[width=0.9\columnwidth]{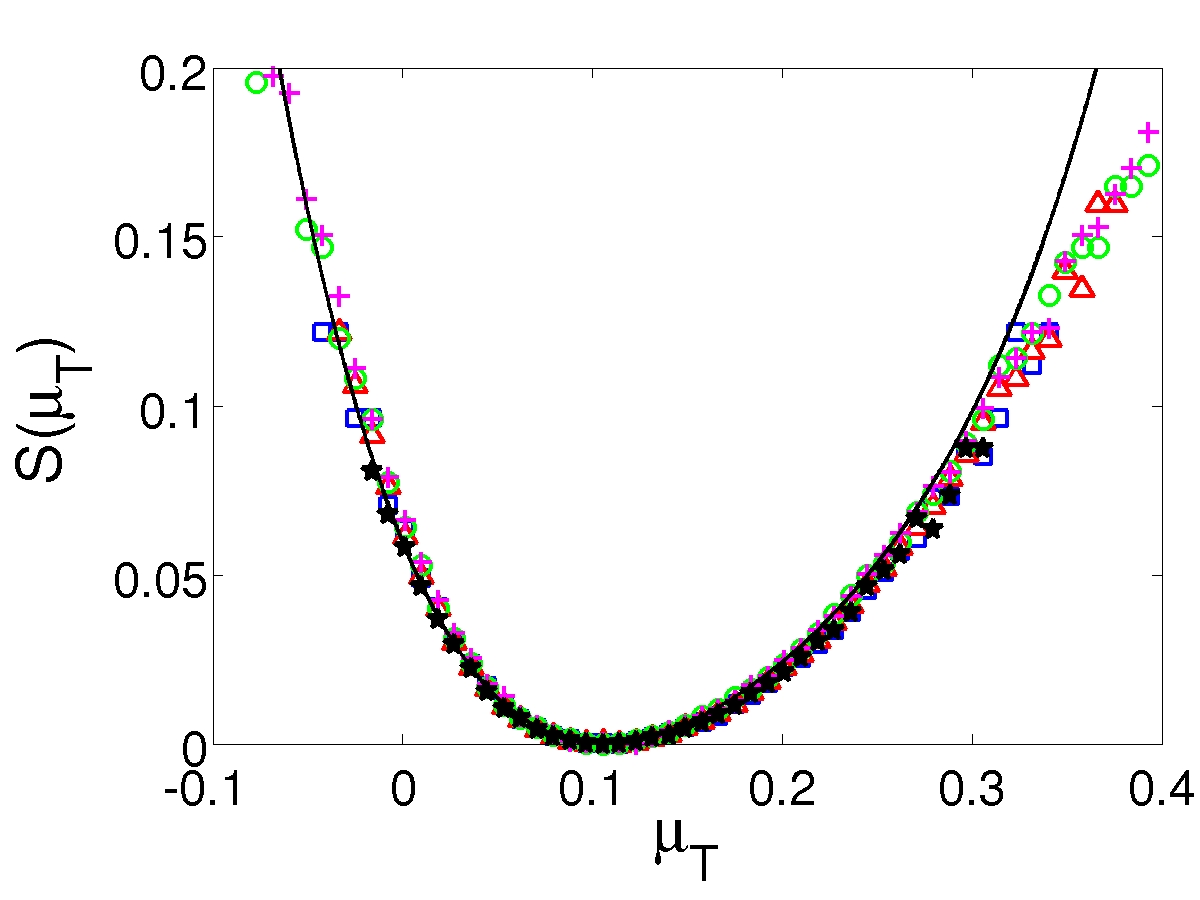}
\caption{\label{fig:cramer}(Color online) The collapse of the Cramer's functions 
$S(\mu)$ versus $\mu$ at $\yplus=62$ for different times $T$, v.i.z,
$T= 35.(+), 45.(\circ), 55.(\vartriangle), 70.(\square)$, and
$100. (\star)$. All times are measured in the units of $\tauL$.
  The continuous line is the polynomial fit as given in 
Eq.~\ref{eq:polynom_cramer} with $\mubar=0.105 [0.088\ \ 0.13]$ and $a_2=3.55[3.09\ \ 4.35]$,
$a_3=-12.60 [-27.48\ \ -4.29]$ and $a_4 = 39.64 [3.84\ \ 90.07]$.
We have used the same polynomial form to fit 
$S(\mu)$ obtained for individual $\yplus$. The maximum and minimum values
of the fitting parameters are given inside square brackets. 
}
\end{center}
\end{figure}

From the PDF of $\muT$ for large $T$ we calculate the Cramer's function 
using Eq.~\ref{eq:cramer}. 
We normalise $P(\muT)$ such that its integral over the range of $\muT$ is unity. 
For $T>T_{\rm exit}$ the Cramer's functions $S(\mu)$ 
calculated at different times $T$ are found to be independent of $T$
as it should be. 
This is shown by the collapse of the Cramer's function calculated at 
different times for a fixed $\yplus = 62$ in Fig.~(\ref{fig:cramer}). 
This proves that the conjecture in \Eq{eq:cramer} hold true. 
Furthermore, the Cramer's function thus found is independent of $\yplus$. 
The Cramer's function has earlier been calculated from DNS of two~\cite{bof+cel+mus03}- and 
three~\cite{bec+bif+bof+cen+mus+tos06} dimensional 
homogenous isotropic turbulence, turbulence in the presence of homogeneous 
shear~\cite{eck+kro+sch02}, and for hydromagnetic 
convection~\cite{kur+bra91}.  This is the first time is has been calculated for a channel flow. 

The connection between the Cramer's function and the PDF of end-to-end
polymer distance was shown in Ref~\cite{bal+fux+leb00} for linear polymers and
in Ref~\cite{che00} for nonlinear polymers. 
We discuss such relations in the next section where it will turn out to be useful
to have an algebraic expression for the Cramer's function.
In the simplest case the Cramer's function is a parabola which implies that the PDF
of FTLEs is a Gaussian distribution. It is clear from Fig.~(\ref{fig:cramer}) that in our
case $S(\mu)$ is not well approximated by a parabola except for $\mu \approx \mubar$. 
The departure from Gaussianity is characterized
by higher (than second) power of $\mu$ in a polynomial expansion of $S(\mu)$. 
The next level of approximation would be to use a fourth--order polynomial for the following two reasons: 
(a) The function $S(\mu)$ in Fig.~(\ref{fig:cramer}) is clearly not symmetric about its
axis, hence we need an odd power of $\mu$ to approximate it. (b) The function $S(\mu)$ must be
convex hence the highest power of $\mu$ appearing in $S(\mu)$ must be even. 
Hence we use fit the fourth-order polynomial,
\begin{equation}
S(\mu) = a_2(\mu -\mubar)^2 + a_3(\mu -\mubar)^3 + a_4(\mu -\mubar)^4 ,
\label{eq:polynom_cramer}
\end{equation}
to our numerical data for $S(\mu)$ averaged over all values of $\yplus$ and 
extract the coefficients $a_2,a_3$, and $a_4$ above. 
To estimate the errors in the coefficients $a_k$ we use the same fit to
$S(\mu)$ obtained for individual $\yplus$ and quote the range of $a_k$ obtained
from such fits as the error in $a_k$.  
The best fit is also plotted in Fig.~(\ref{fig:cramer}). The coefficients
corresponding to the best fit and their errors are given in the caption
of Fig.~(\ref{fig:cramer}).
%
\subsection{Statistics of polymer extensions: Oldroyd-B model}
\label{subsec:PDF}
Before we present detailed results on statistics of polymer
extension let us precisely define the Weissenberg number, $\Wi$. 
In simulations the Weissenberg number is defined as the
ratio of the characteristic time-scale of the polymer, $\taup$
over a characteristic time scale of the fluid.  
Different definitions of the characteristic time scale for fluid has been used in 
literature to define the Weissenberg number. 
Refs.~\cite{per+mit+pan06,jin+col07,wat+got10} use the Kolmogorov time scale 
$\tau_{\eta}$ to define the Weissenberg number. We denote this
Weissenberg number by $\Wi_{\eta} = \taup/\tau_{\eta}$ where
$\tau_{\eta}$ is the Kolmogorov time scale. 
In this paper we principally use the following definition
for Weissenberg number
\begin{equation}
\Wi \equiv \mubar \taup
\end{equation} 
where $\mubar$ is the location of the minima of the Cramer's
function $S(\mu)$. Our choice has two principal advantages. 
Firstly in channel flows the Kolmogorov scale depends on the
wall-normal coordinate and hence is not unique. Secondly
and more importantly a proper choice of Weissenberg number
gives the coil-stretch transition of the polymer
at $\Wi \approx 1$ which is exactly what we obtain. 
To compare with earlier simulations, which were all done in homogeneous
flows, we also calculate $\Wi^{\rm wall}_{\eta}$ and $\Wi^{\rm center}_{\eta}$, 
where we use the Kolmogorov time scale at the wall and at the center of 
the flow respectively. We typically obtain, $\Wi^{\rm wall}_\eta \approx 30 \Wi$ and 
 $\Wi^{\rm center}_{\eta}\approx 5 \Wi$. The different values
of $\Wi$ that we use are given below, in parentheses we give the corresponding
values of $\Wi^{\rm center}_{\eta}$ for easy comparison with 
earlier simulations of homogeneous and isotropic turbulence. 
For the Oldroyd-B model, 
$\Wi(\Wi^{\rm center}_{\eta})= 0.1(0.5)$, $0.2 (1)$,$0.3(1.5)$, and $0.5(2.5)$ 
and 
$\Wi(\Wi^{\rm center}_{\eta}) = 0.1 (0.5)$, $0.3 (1.5)$,$0.5 (2.5)$,$1.5 (7.5)$, 
$2.5 (12.5)$,$3.5 (17.5)$,$4.5 (22.5)$,$5.5 (27.5)$,$7 (35)$, and $10 (50)$ 
for the FENE model.
We use $R_0=10^{-7}$, $10^{-8}$, and
$R_{\rm max}/R_0 = 100$ and $1000$ for the FENE model. 

\begin{figure}[!h]
\begin{center}
\includegraphics[width=0.9\columnwidth]{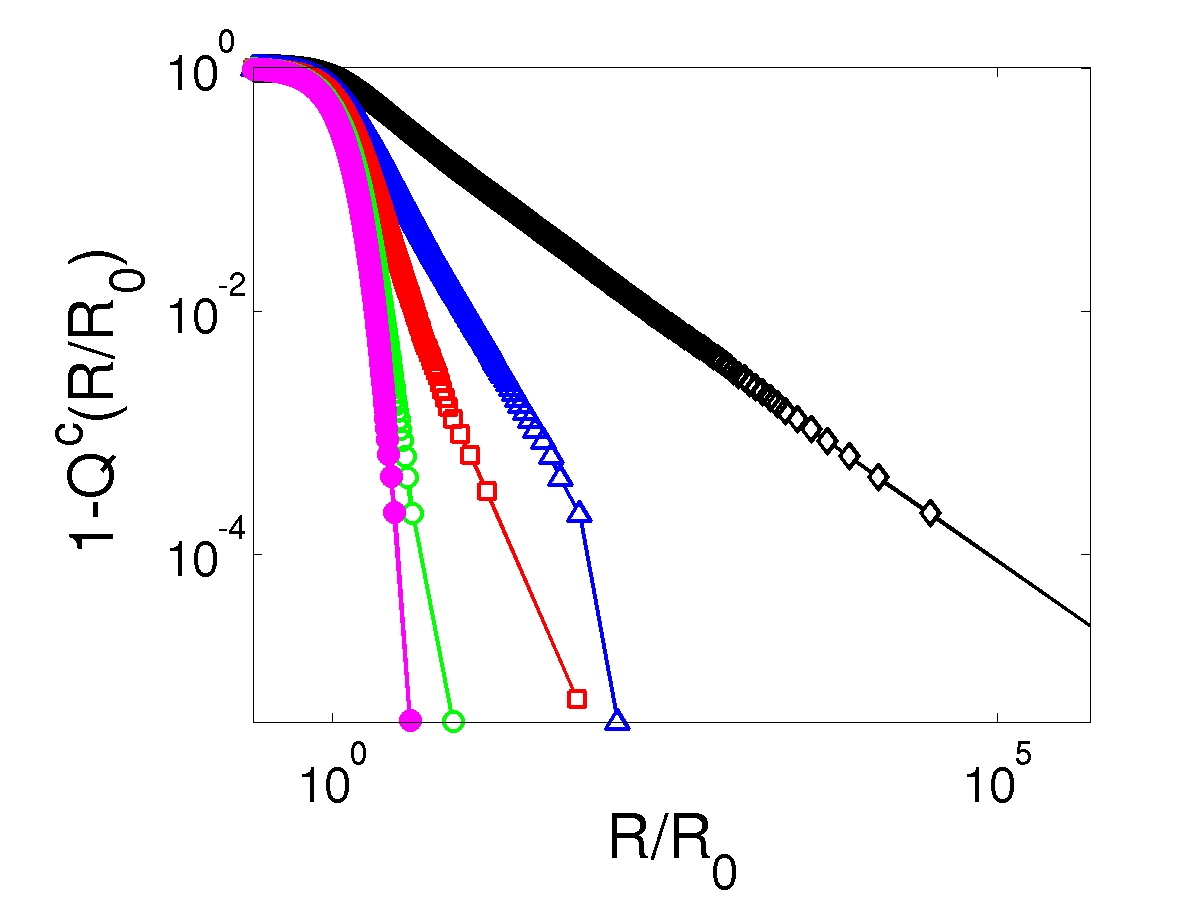}
\caption{\label{fig:cpdf_taup}(Color online) 
Log-log plot of the cumuliative PDF $Q^c(R)$ of the polymer extensions $R$ 
as a function of $R$ for different values of $\Wi$, v.i.z, 
$\Wi= 0.05($\textbullet$), 0.1(\circ), 0.2(\square), 0.3(\vartriangle)$, and $0.5(\lozenge)$.  }
\end{center}
\end{figure}

\begin{figure}[!h]
\begin{center}
\includegraphics[width=0.9\columnwidth]{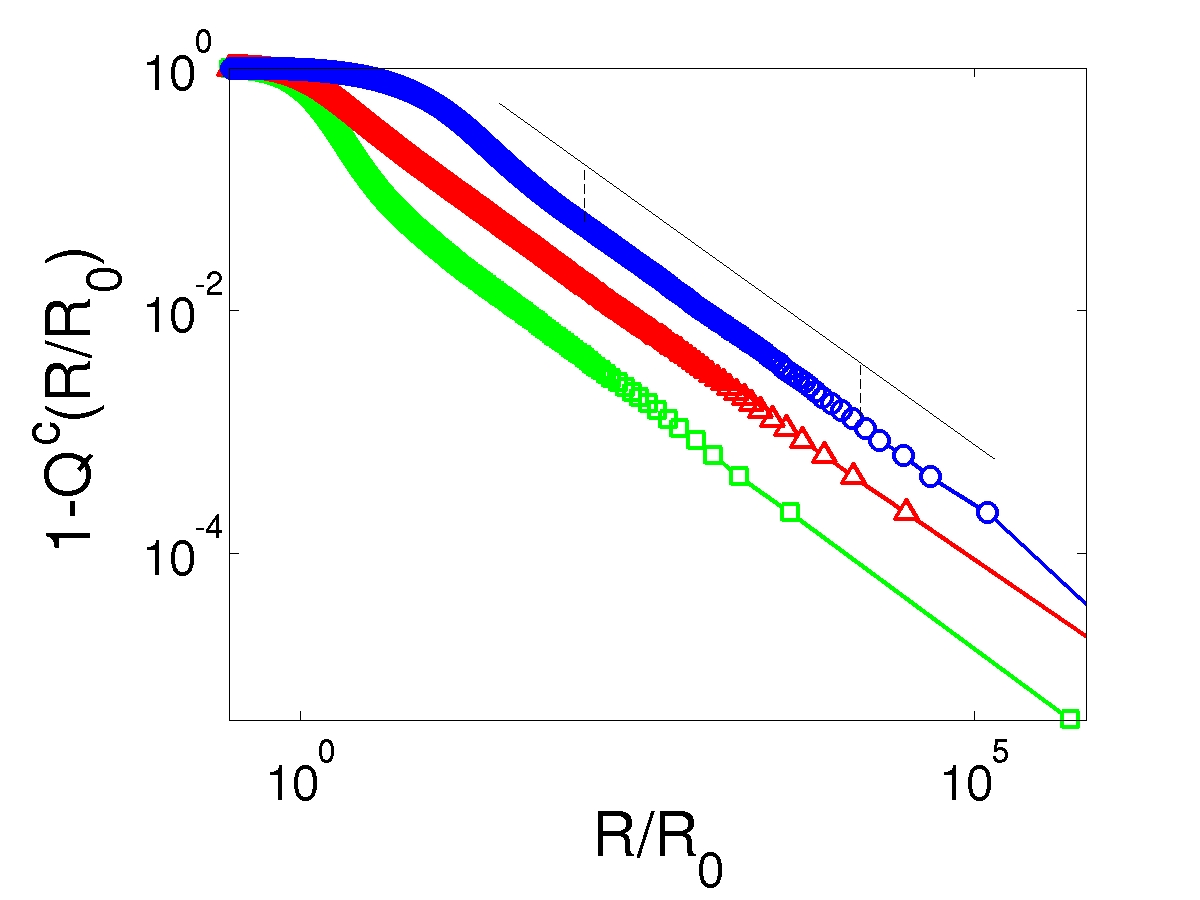}
\caption{\label{fig:cpdf_yplus}(Color online) 
Log-log plot of the cumuliative PDF $Q^c(R)$ of the polymer extensions $R$ 
as a function of $R$ for different $\yplus$, v.i.z,
$\yplus= 8(\circ), 74(\vartriangle)$, and
$180 (\square)$. We fit a straight line to the data between the
two dashed vertical lines to calculate $\alpha$. This fit is
shown as the black line. }
\end{center}
\end{figure}

Let us first present the results for the Oldyroyd-B model.
Here we expect to see a power-law behavior for the PDF of polymer
extensions, $Q(R) \sim R^{-\alpha-1}$~\cite{bal+fux+leb00} for large $R$.  
In general, the calculation of PDFs from numerical data is 
plagued by errors originating from the binning of the data to make 
histograms. 
Thus it is often a difficult task to extract exponents such as 
$\alpha$ from such PDFs. 
A reliable estimate of such an  exponent can be obtained by using 
the rank-order method~\cite{mit05a} to calculate the corresponding 
cumulative probability distribution function,
\begin{equation} 
Q^c(R) \equiv \int_0^{R}Q(\xi) d\xi
\label{qr}
\end{equation}
If the PDF has a scaling range the cumulative PDF also shows scaling, i.e.,  
$Q^c(R) \sim  R^{-\alpha}$.
These cumulative PDFs are plotted in Fig.~(\ref{fig:cpdf_taup}) for different values of $\Wi$ at 
fixed wall distance  $\yplus=74$
The cleanest
power-law is seen for $\Wi = 0.5$. So we choose this 
Weissenberg number for further detailed investigation. First
we show that the exponent of the power-law ($\Wi=0.5$) 
$\alpha=0.81\pm0.02$ does not depend on the $\yplus$ although
the range over which scaling is obtained does, Fig.~(\ref{fig:cpdf_yplus}).
The exponent $\alpha$ is obtained 
by fitting a power-law for five different values of $y^+$. The mean is reported 
as the exponent above and the standard deviation from the mean is reported 
as the error.  
 
This exponent $\alpha$ can be obtained from the Cramer's function
$S(\mu)$ using the set of couple equations Eq.~(\ref{eq:alpha}) and 
(\ref{eq:beta})~\cite{bal+fux+leb00} which we rewrite below,
\begin{equation}
\alpha = S^{\prime}(\beta+\frac{1}{\taup}-\mubar)
\end{equation}
where $\beta$ must be obtained by solving the differential equation,
\begin{equation}
S(\beta+\frac{1}{\taup}-\mubar) -\beta S^{\prime}(\beta+\frac{1}{\taup}-\mubar) =0
\end{equation}
Had the Cramer's function been well approximate by a parabola of the
form $S(\mu) = (\mu-\mubar)^2/\Delta$, Eq.~(\ref{eq:alpha}) would
simplify to $\alpha = (2/\Delta)(1/\taup-\mubar)$. We have
checked that this quadratic approximation does not give accurate
result for $\alpha$ in our case. Using the algebraic expression
for $S$ given in Eq.~\ref{eq:polynom_cramer}, we numerically
solve Eqs.~(\ref{eq:alpha}) and (\ref{eq:beta}). This give
$\alpha = 0.9 \pm 0.29$ which agrees with the results obtained from the 
cumulative PDF of polymers within error bars. 
We note here that the $\alpha$ we calculate using the Cramer's function
has large margin of error because the $\alpha$
depends sensitively on the coefficients $a_k$ in 
Eq.~\ref{eq:polynom_cramer}.
To find these coefficients accurately we need to know the Cramer's function
accurately for a large range of its argument not just the location of its
minima. 
Numerically this is a difficult task and would require collecting data 
over very long times.

\begin{figure}[!h]
\begin{center}
\includegraphics[width=0.9\columnwidth]{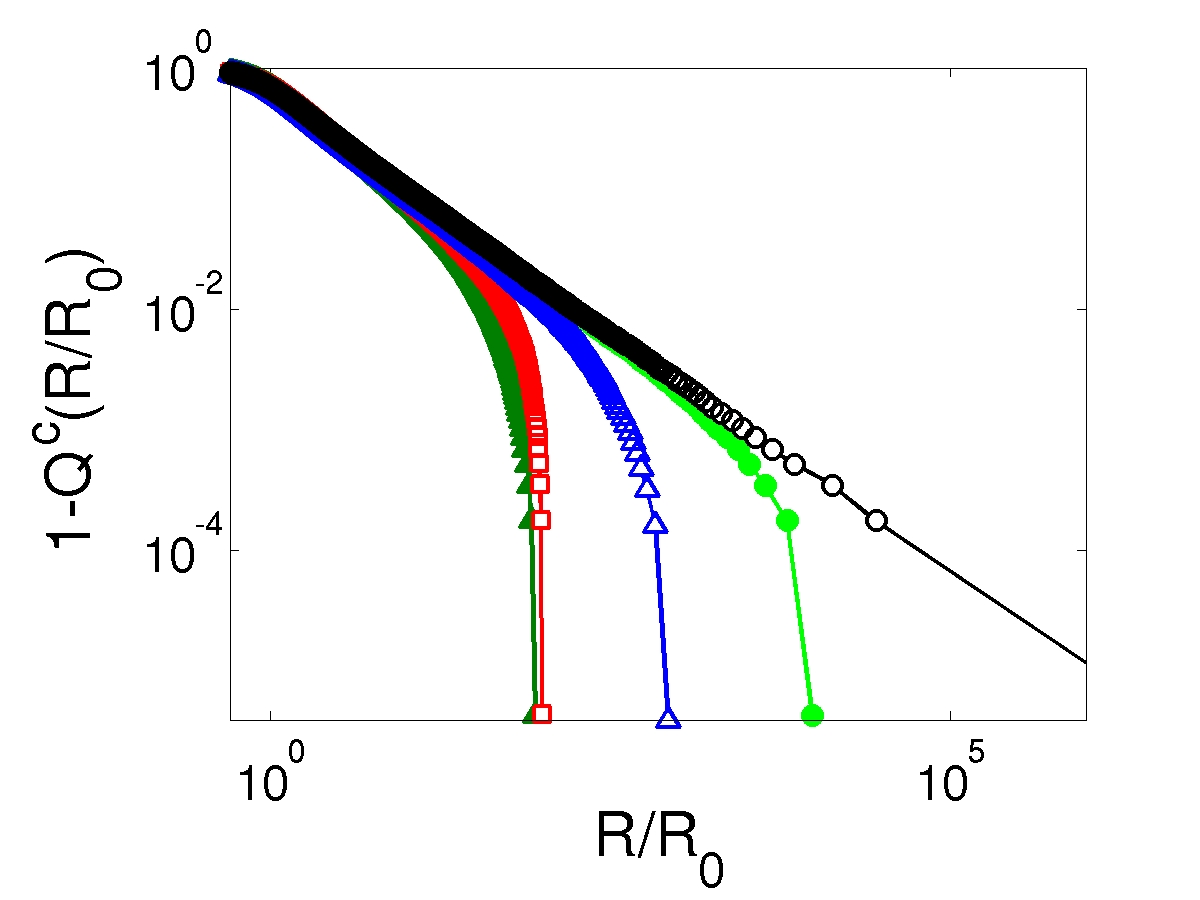}
\caption{\label{fig:cpdf_cutoff}(Color online) 
The cumulative PDF $Q^c(R)$ of the polymer extensions $R$ 
as a function of $R$ for the Oldroyd-B model ($\circ$), Oldroyd-B model
with all polymers with $R/R_{\rm max} > R_{\rm cutoff}$ removed 
with $R_{\rm cutoff} = 10^4 ($\textbullet$)$ and $R_{\rm cutoff} = 100 (\square)$,
FENE model with $R_{\rm max}/R_0 = 10^3 (\vartriangle)$ and with
$R_{\rm max}/R_0 = 10^2 (\blacktriangle)$.
}
\end{center}
\end{figure}

Finally let us comment on the possible experimental determination of the
exponent $\alpha$. In practice no polymers are linear and in most cases
the ratio of $R_{\rm max}$ (maximum possible extension of the polymer) 
to  $R_0$ (the equilibrium length) ranges between 
$100$ and $1000$. To see the effect of a maximum extension, we first select 
one of the cumulative PDFs plotted in Fig.~(\ref{fig:cpdf_yplus}), say for $\yplus=74$.   
From this cumulative PDF we remove all the polymers for which $R$ is so 
large that $R/R_0 > R_{\rm cutoff}$ where we choose 
$R_{\rm cutoff} = 100$ and $1000$.
The resultant cumulative PDFs are plotted in Fig.~(\ref{fig:cpdf_cutoff}) 
where the original cumulative PDF is also plotted for comparison. 
It can be seen that the scaling
behavior, although present, is valid over a much smaller range. 
In the same figure we have also plotted the cumulative PDF for the FENE 
model with $R_{\rm max}/R_0 = 1000$. This also shows scaling with a reduced 
range. Thus we expect that in experiments similarities to this scaling law 
should be visible although it may be difficult to detect because of a reduced 
range of scaling. 
\subsection{Statistics of polymer extensions: FENE model}\label{sub:FENE}

So far we have described the polymer statistics for $\Wi \leq 0.5$.
As we increase the $\Wi$ and make it close to unity no stationary statistics
of the polymers is obtained. We interpret this by noting that we are close
to the coil-stretch transition. A stationary state can be obtained 
either by including the feedback from the polymers into the fluid or
by using nonlinear polymers e.g., the FENE model. We choose the second
option. In the FENE model we have used $R_{\rm max}/R_0 = 100$ and $1000$. 
Our results as reported below does not depend on this parameter. 

\begin{figure}[!h]
\begin{center}
\includegraphics[width=0.9\columnwidth]{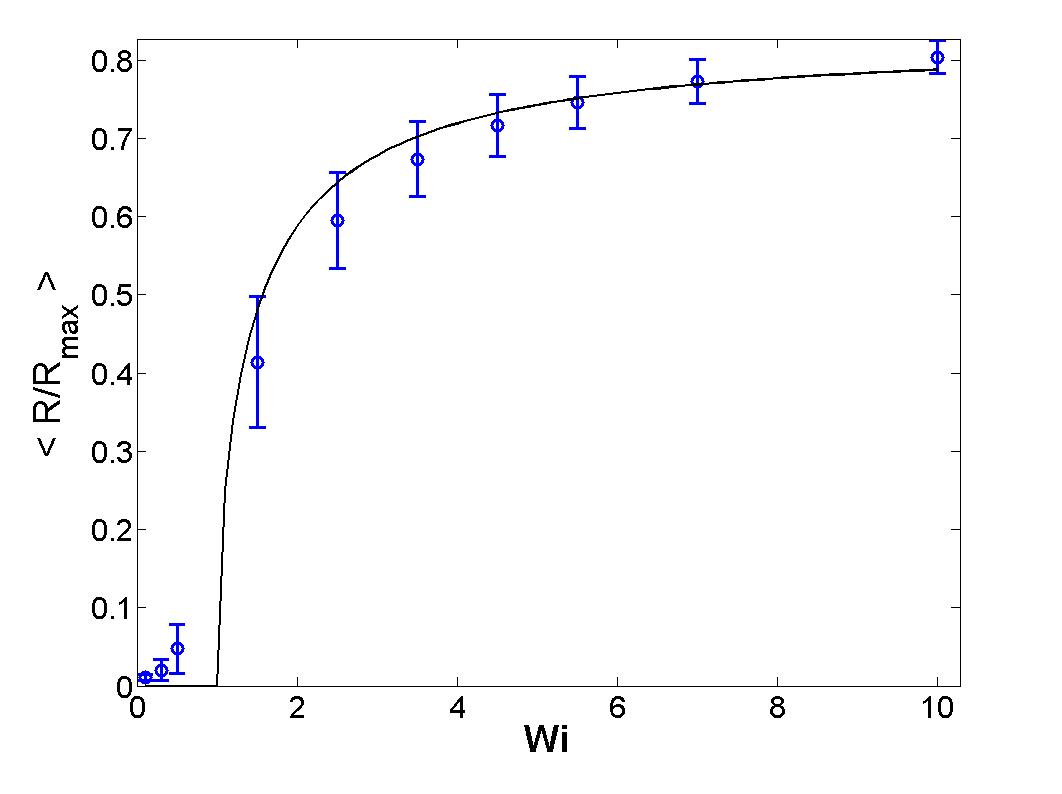}
\caption{\label{fig:chertkov}(Color online)
The mean of normalized polymer extensions $\bra{ R/\Rmax }$ 
as a function of Weissenberg number $\Wi$. The mean is calculated
over the whole channel and the standard deviation is shown
as error bar.  The continuous line is the right hand side of \eq{eq:che} calculated
for $\Wi >1$.
}
\end{center}
\end{figure}

Let us first consider the mean extensions of the polymers averaged over the
whole channel as a function of the Weissenberg number. 
Using a saddle point approximation Chertkov~\cite{che00} has shown that for  
$\Wi > 1$ the mean polymer extension  obeys the implicit relation
\begin{equation}
\bra{\frac{R}{\Rmax}} = -\frac{1}{\mubar}f(\bra{R})
\label{eq:che}
\end{equation}
where $f$ is the FENE force. 
In Fig~\ref{fig:chertkov} we show that we obtain reasonable agreement 
between between this analytical prediction and our numerical results
for different values of the Wisenberg number. 
The error-bars in this plot are the variance of the polymer extension calculated
over the channnel. 

\begin{figure}[!h]
\begin{center}
\includegraphics[width=0.9\columnwidth]{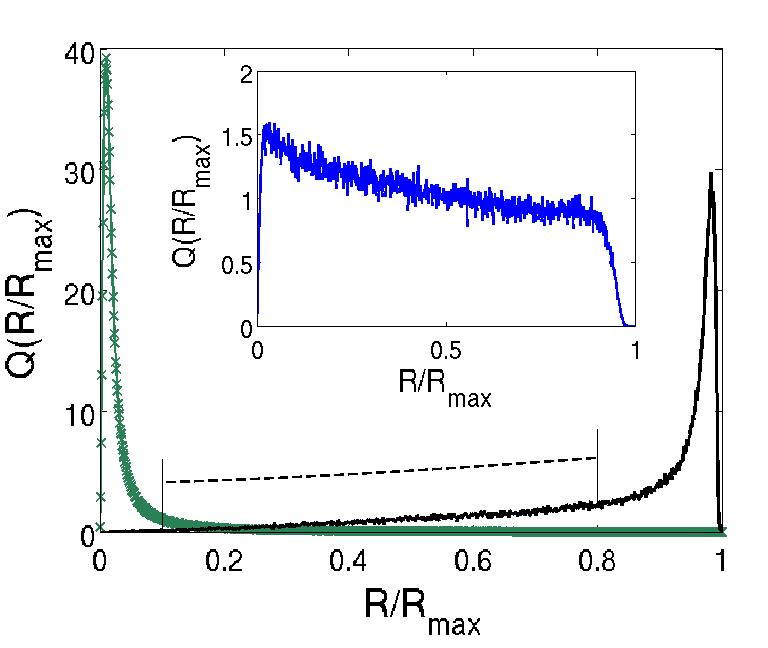}
\caption{\label{fig:polypdf1}(Color online) 
The PDF of polymer extensions $Q(R)$ as a function of $R$
for different $\Wi$ showing the coil-stretch transition. 
The line with $(\times)$ symbols is for $\Wi=0.5$ ($\taup=5$),
the continuous line is for $\Wi=10$ ($\taup=100$) and
the inset is for $\Wi=1.5$ ($\taup=15$). The PDF for
$\Wi=10$ is multiplied by $2$ to make it clear in the
same scale. The dashed line shows power-law scaling 
with exponent $\alpha =1.48 $.  
}
\end{center}
\end{figure}

\begin{figure}[!h]
\begin{center}
\includegraphics[width=0.9\columnwidth]{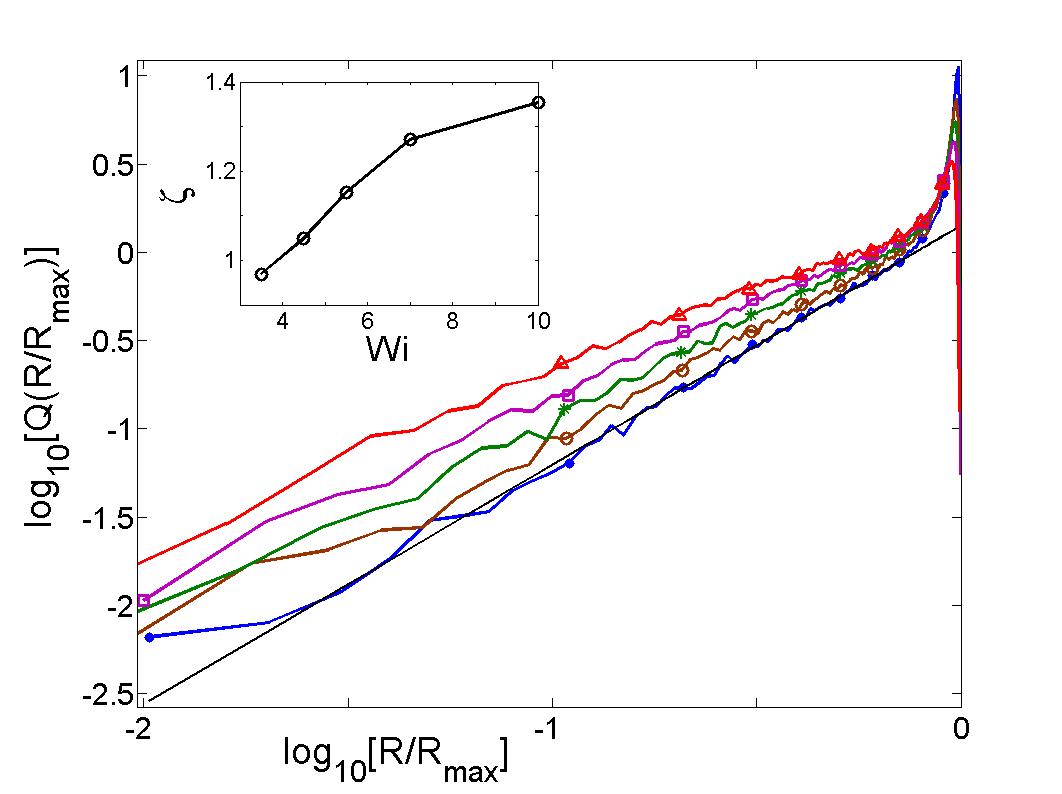}
\caption{\label{fig:slope_pdf}(Color online) Log-log plot of the PDF of
polymer extensions for  $\Wi = 3.5 (\vartriangle), 4.5 (\square),5.5 (+),7(\circ),
10 (\bullet)$.  The straight line is a fit to the PDF for $\Wi = 10$. 
Similar fits yield the exponents $\zeta$ which are plotted as a function of $\Wi$
in the inset. }
\end{center}
\end{figure}

Let us now consider the full PDF of the polymer extension. 
In Fig.~(\ref{fig:polypdf1}) we plot the PDF for three different values of 
the Weissenburg number, $\Wi = 0.5$, $1.5$ and $10$.
The coil-stretch transition is clearly demonstrated in this figure. 
For $\Wi = 0.5$ the PDF is peaked near zero which corresponds to the
coiled state. 
For $\Wi=1.5$ the peak of the PDF is still close
to zero but the PDF is well spread over the whole range. 
At  $\Wi=10$ the PDF has a peak near $R_{\rm max}$;  this is the 
stretched state of  the polymer.  
In this Figure we have plotted the PDFs for $\yplus=74$.  
The PDF at other wall-normal coordinates in the channel shows the same 
qualitative nature.
Similar plots of the PDF of polymer extensions but for a simple model of 
polymers in uniform shear has been obtained in Ref.~\cite{che+kol+leb+tur05}.
A more careful scrutiny, however, reveals differences between our results and that  
of Ref.~\cite{che+kol+leb+tur05}  for $\Wi = 10$. 
In particular,  we do not observe the plateau in the PDF seen in Fig 2 of 
Ref.~\cite{che+kol+leb+tur05}.
However, it is possible to observe a power-law behavior of the left-tail of the
PDF as shown in  Fig.~\ref{fig:slope_pdf}. 
Plots of the PDF of polymer extensions have also been recently obtained in
experiments~\cite{liu+ste10}. 
For strong shear the experiments results have qualitative agreement with 
the results of Ref.~\cite{che+kol+leb+tur05} including the presence of
the plateau, although quantitative agreement is still lacking.
The disagreements of our results with that of Ref.~\cite{che+kol+leb+tur05}
might be due to spatial inhomogeineity of channel flow compared to the
case of uniform shear. 

\begin{figure}[!h]
\begin{center}
\includegraphics[width=0.9\columnwidth]{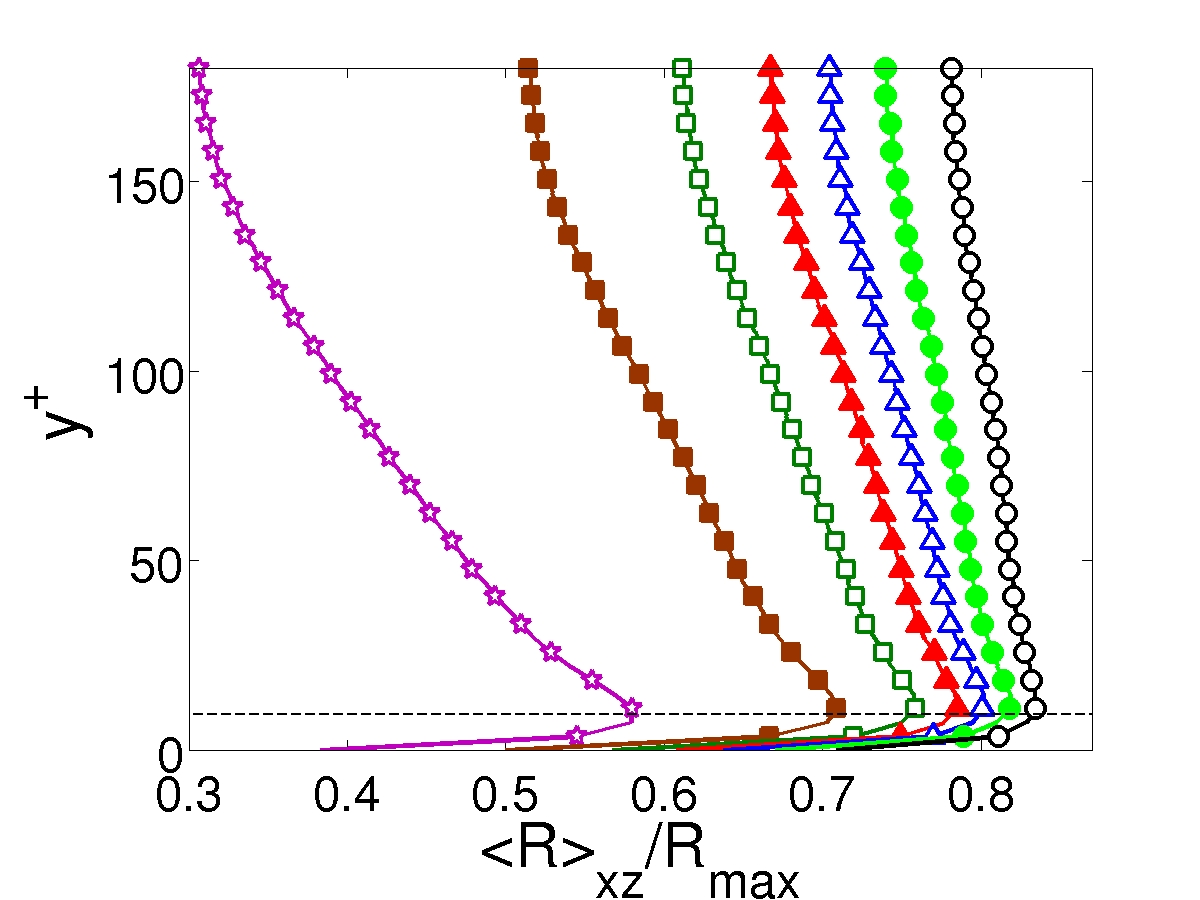}
\caption{\label{fig:Rmean_yplus}(Color online) 
The average polymer extensions $\bra{R}_{xz}$ as a function 
of the wall-normal coordinate $\yplus$ for different
Weissenberg numbers, v.i.z., $\Wi = 1.5 (\star),2.5 (\blacksquare),
3.5 (\square), 4.5 (\blacktriangle), 5.5 (\vartriangle), 7 ($\textbullet$), 10 (\circ). $
The maximum occurs at $\yplus\approx 10$.}
\end{center}
\end{figure}

\begin{figure}[!h]
\begin{center}
\includegraphics[width=0.9\columnwidth]{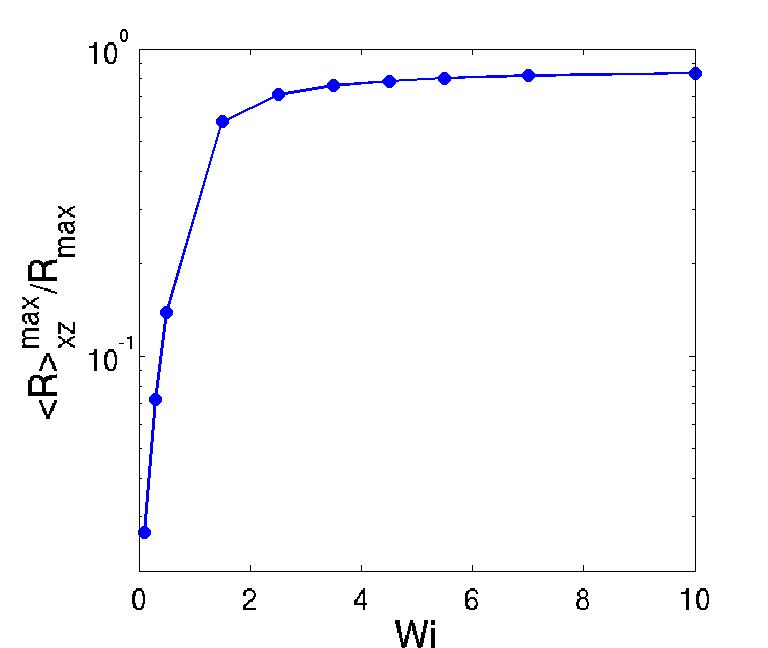}
\caption{\label{fig:max_vs_Wi}(Color online) 
The maxima of the average polymer extensions $\bra{R}_{xz}$
across the channel  as a function 
of $\Wi$ in semilog scale. 
}
\end{center}
\end{figure}

The effects of spatial inhomogeneity is also seen in 
Fig.~(\ref{fig:Rmean_yplus}), where we show how the mean polymer extension 
$\bra{R}_{xz}$, where the averaging is over the stream-wise and the 
span-wise direction, changes with $\Wi$ across the 
channel for $\Rmax/R_0 = 100$. 
For a given $\Wi$ the average polymer extension is small near the wall,
increases to a maximum around  $y^+\approx 10$ (this corresponds to the region of 
maximum strain), and then decreases towards 
the center of the domain where the flow is close to homogeneous turbulence.
A similar trend is also seen for $\Rmax/R_0 = 1000$.
This trend has been seen in earlier DNS of polymeric turbulence
in channel flows~\cite[see e.g.,][and references therein]{whi+mun08}.  
Note however that for larger values of $\Wi$ the average polymer extension becomes 
almost uniform across the channel (except very near the wall
where it is always small).
This is because the polymers that are stretched close to the wall on reaching the 
centerline are not able to relax fast enough because the polymer relaxation time scales 
are much larger than the fluid time scales.
The maximum extension increases as a function of Weisenberg number for small Weisenberg numbers and saturates
for higher values, see Fig. \ref{fig:max_vs_Wi}.
\subsection{Statistics of polymer orientation}
\label{subsec:angle}
\begin{figure}
\begin{center}
\includegraphics[width=0.9\columnwidth]{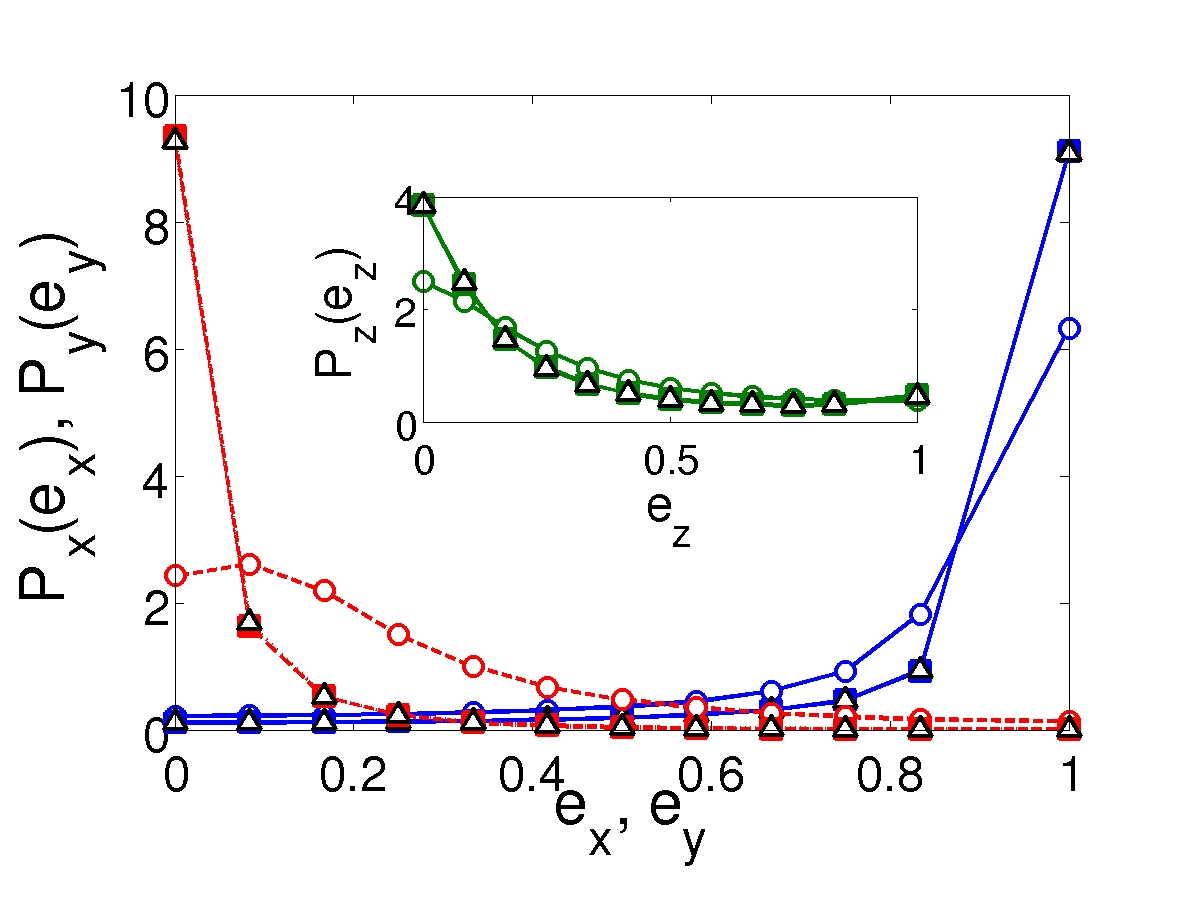}
\put(-100,135) {\Large{(a)}} \\
\includegraphics[width=0.9\columnwidth]{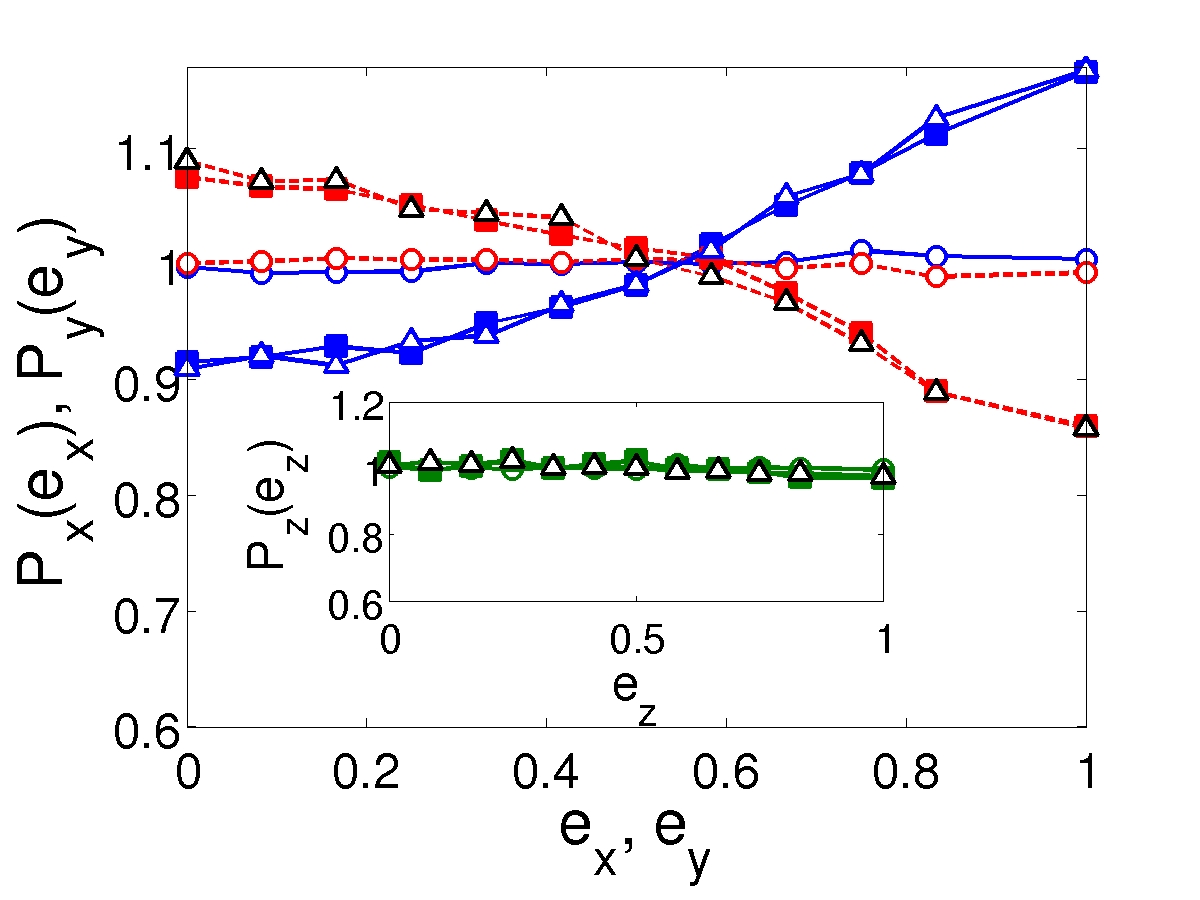}
\put(-100,135) {\Large{(b)}} \\
\caption{\label{fig:ehatwall}(Color online) 
PDF of the three direction
cosines of polymer end-to-end separation vector ${\bm R}$ 
(a) for polymers near the wall  and
(b) for polymers at the center of the channel.
Three different values of $\Wi$ are used. Namely,
$\Wi=0.1 (\circ), 1.5 (\vartriangle), 4.5 (\blacksquare)$.
The data for $\Wi = 1.5$ and $4.5$ coincide on each other. 
The PDFs of $e_x$ and $e_y$  are respectively 
plotted using \textcolor{blue}{continuous line with symbols} ($P_{\rm x}$) and 
\textcolor{red}{dashed lines with symbols} ($P_{\rm y}$). 
The inset shows the PDF of $e_z$, $P_{\rm z}$.
}
\end{center}
\end{figure}

In this section we present the results related to the orientation of the 
polymers.  First let us discuss the orientation of the polymers with respect
to the geometry of the channel. Let us denote the unit vector along  
${\bm R}$ to be $\hate$.  The PDF of the three components of $\hate$, 
$e_x$, $e_y$ and $e_z$, (i.e., three direction cosines of ${\bm R}$) are 
plotted in Fig.~(\ref{fig:ehatwall}a) for polymers close to the wall 
($\yplus\approx 7$) and for three different values of $\Wi$. 
For $\Wi < 1$ , i.e., below the coil-stretch transition the polymers
are almost equally probable to point in any direction or in other words
as the polymers are coiled as a sphere no preferential direction is selected. 
Above the coil-stretch transition polymers close to the wall have a high probability 
of being oriented along the $x$ axis, which is the stream-wise direction. 
This trend has been observed earlier in Ref.~\cite{gup+sur+kho04}. 
A similar plot for polymers close to the center line ($\yplus\approx 180$)
is given in Fig.~(\ref{fig:ehatwall}b).
 For small $\Wi$ all
directions are equally probable. But as $\Wi$ increases
here too the polymers get preferentially oriented along the stream-wise 
direction although the trend is much weaker than near the wall. 

\begin{figure}
\begin{center}
\includegraphics[width=0.9\columnwidth]{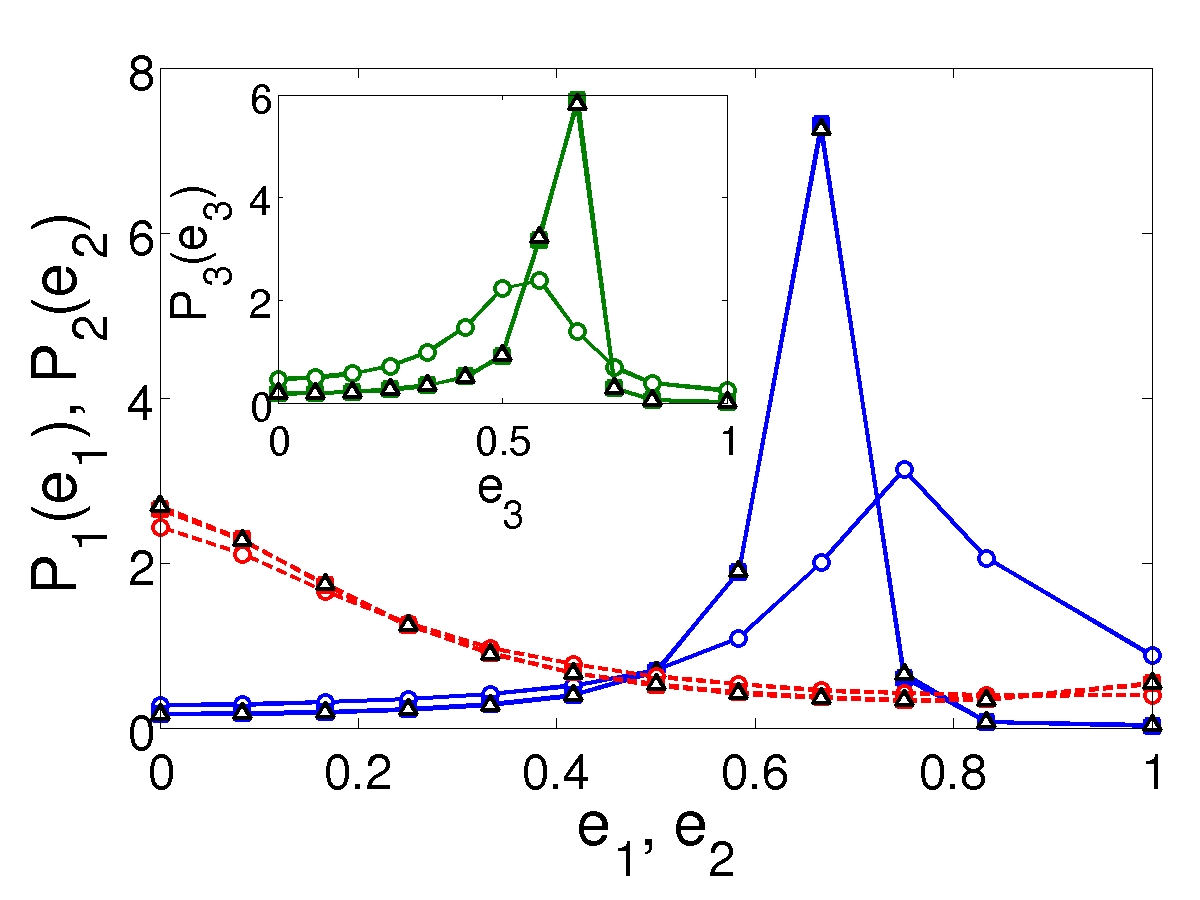}
\put(-40,120) {\Large{(a)}} \\
\includegraphics[width=0.9\columnwidth]{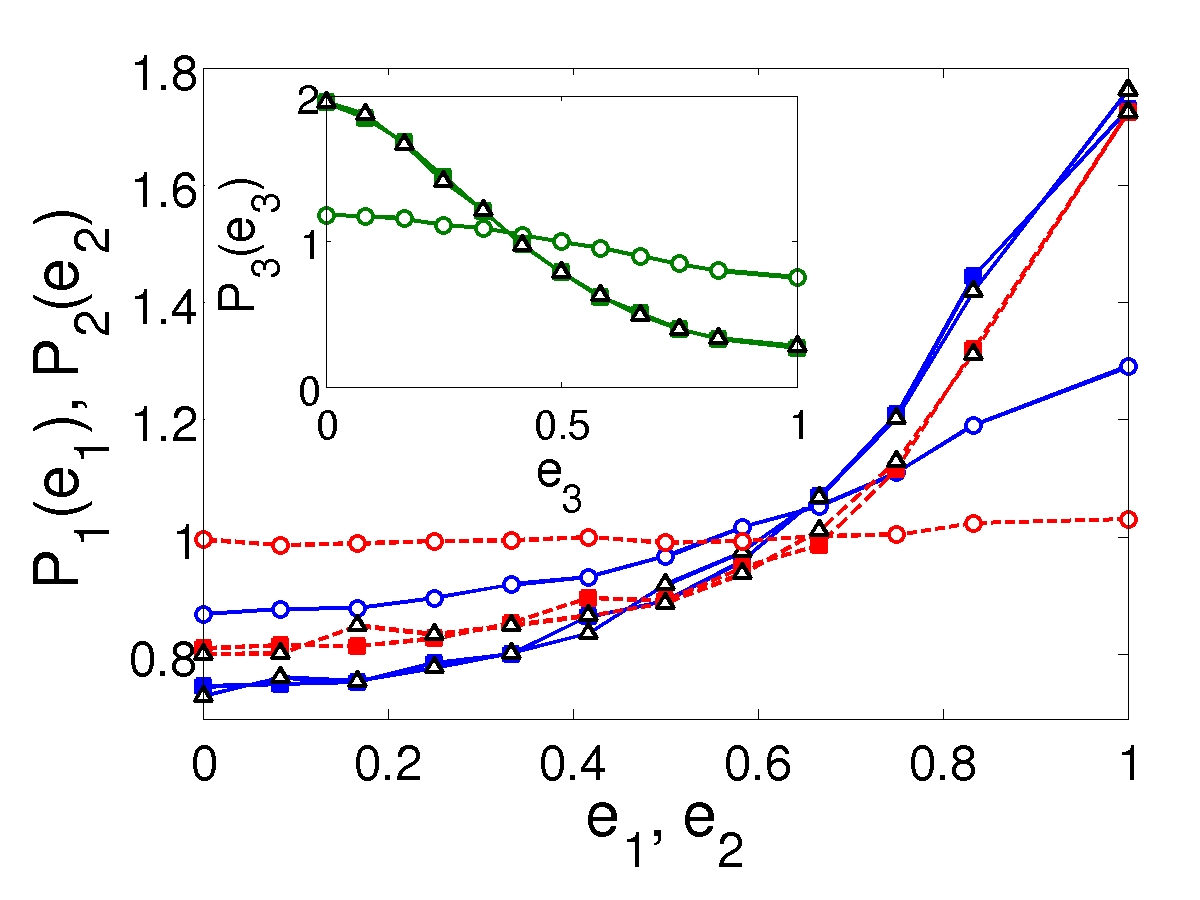}
\put(-40,40) {\Large{(b)}} 
\caption{\label{fig:ehatstrain_wall}(Color online) 
PDF of $e_1$, $e_2$ and $e_3$, components of
the unit vector along ${\bm R}$ along the three principal
directions of strain: (a) for polymers near the wall, 
and (b) for polymers near the centerline.
Three different $\Wi$ are used. Namely,
$\Wi=0.1 (\circ), 1.5 (\vartriangle), 4.5 (\blacksquare)$.
The data for $\Wi = 1.5$ and $4.5$ coincide on each other. 
The PDFs of $e_1$ and $e_2$  are respectively 
plotted using \textcolor{blue}{continuous line with symbols} ($P_{\rm 1}$) and 
\textcolor{red}{dashed line with symbols} ($P_{\rm 2}$). 
The inset shows the PDF of $e_3$, $P_{\rm 3}$.
}
\end{center}
\end{figure}

We have also investigated the orientation of the polymers with respect to the
three principal directions of the rate of strain tensor. 
For this purpose we first determine the three real eigenvalues of the 
symmetric rate of strain tensor and order them such that
$\lambda_1 > \lambda_2 > \lambda_3$. 
We denote the components of the unit vector $\hate$ (which is the unit 
vector along ${\bm R}$) along these three 
perpendicular directions by $e_1$, $e_2$ and $e_3$; these
are merely the cosines of the angles between ${\bm R}$ and the three
principal directions of the strain tensor. 
The PDFs of $e_1$, $e_2$ and $e_3$  are plotted in 
the Fig.~(\ref{fig:ehatstrain_wall}a)
for polymers close to the wall ($\yplus\approx 7$) and for three different 
values of $\Wi$. 
The peak seen in Fig.~(\ref{fig:ehatstrain_wall}) 
corresponds to the polymers orientating along the stream-wise direction as 
shown already in Fig.~(\ref{fig:ehatwall}). 
Interestingly the polymers are not preferentially
oriented along the strongest direction of strain $\lambda_1$ but along the
stream-wise direction. 
This has an angle of about 45 degrees with respect to the x-axis since the
main component of the strain rate comes from the wall-normal shear $\partial U/ \partial y$.

Close to the centerline however the PDFs look quite 
different [Fig.~(\ref{fig:ehatstrain_wall}b)]. 
For small $\Wi$ there is no preferential orientation but as $\Wi$ increases
the polymers develops a trend of orienting parallel to the direction
of either $\lambda_1$ or $\lambda_2$ and shows anti-alignment to $\lambda_3$.  

\begin{figure}
\begin{center}
\includegraphics[width=0.9\columnwidth]{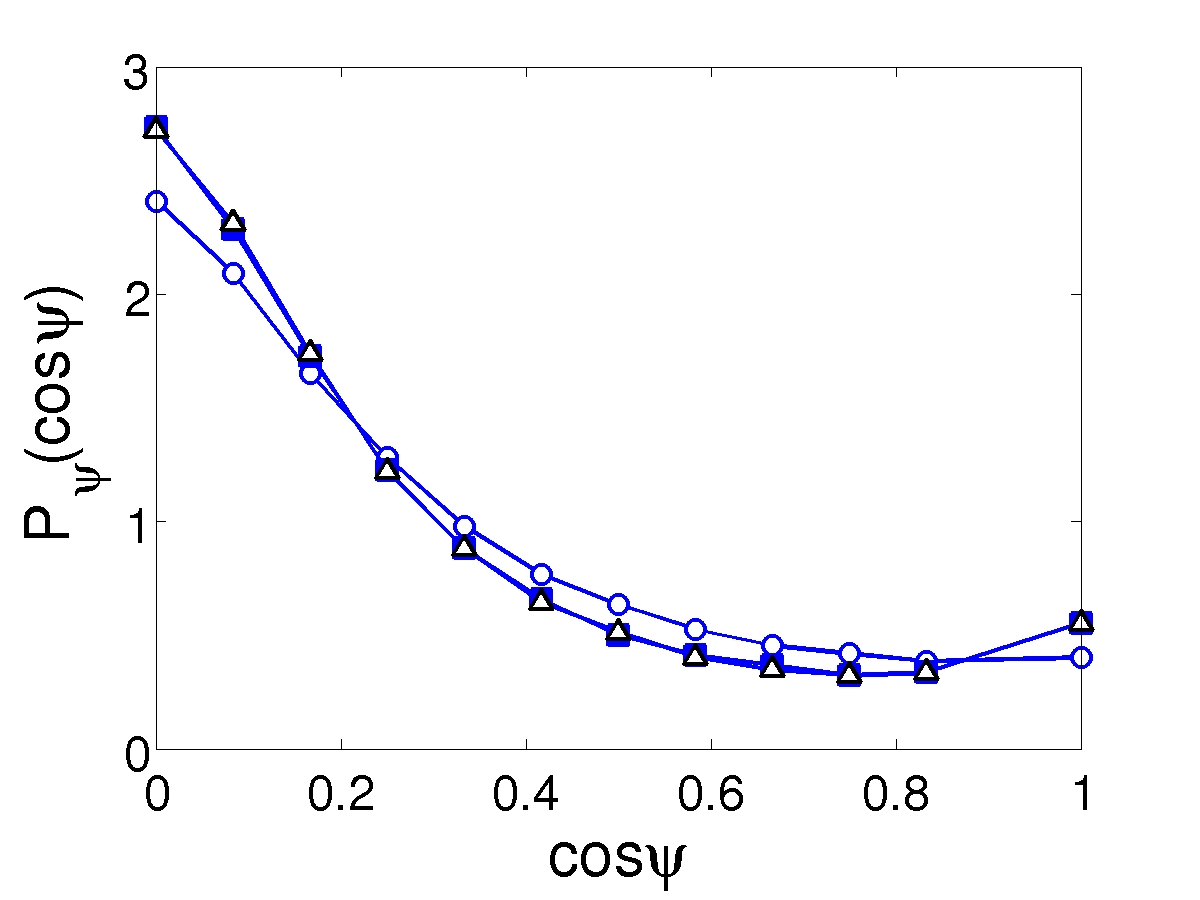}
\put(-100,100) {\Large{(a)}} \\
\includegraphics[width=0.9\columnwidth]{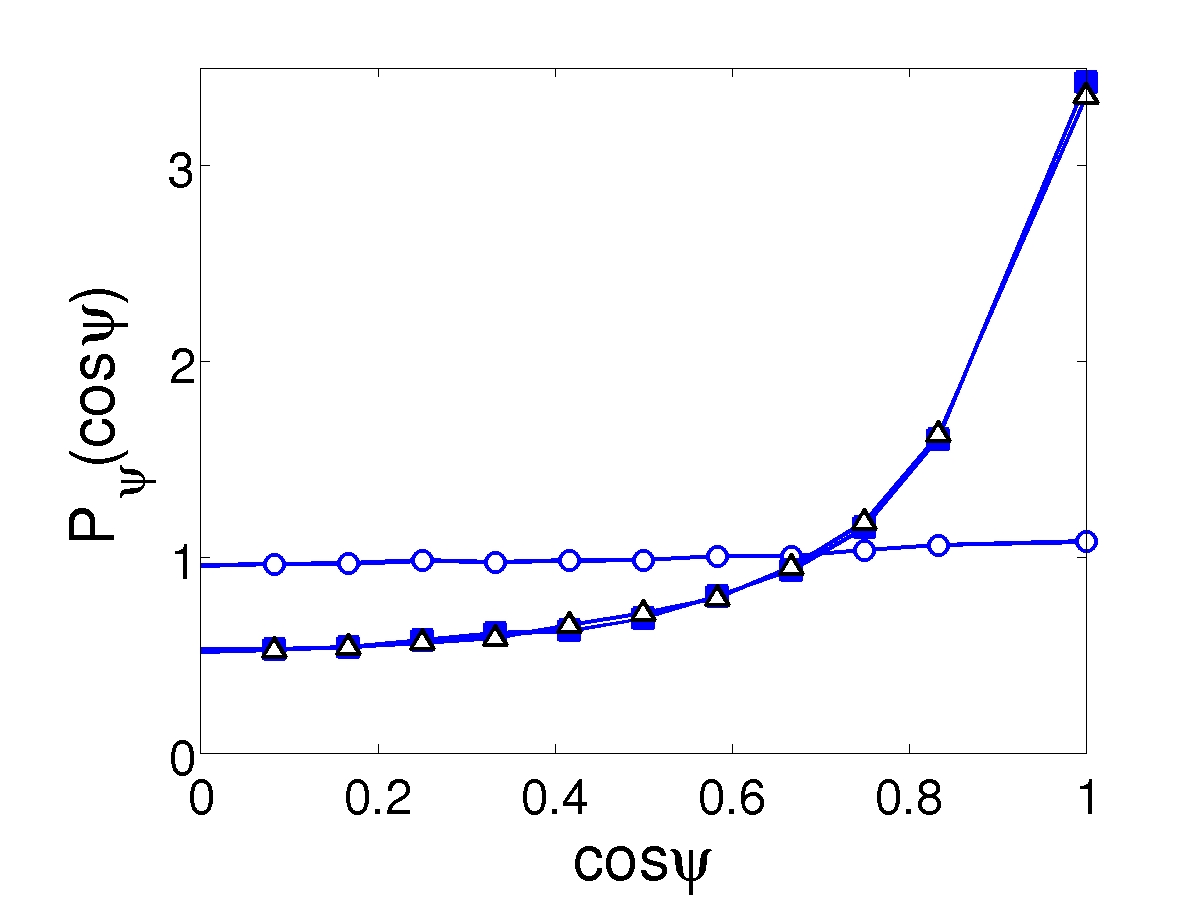}
\put(-100,100) {\Large{(b)}} \\
\caption{\label{fig:psi}(Color online) 
PDF of $\cos(\psi)$, where $\psi$ is the angle between the polymer 
end-to-end vector ${\bf R}$ and vorticity,
(a) for polymers near the wall and
(b) for polymers near the centerline. 
Three different valus of $\Wi$ are plotted. Namely
$\Wi=0.1 (\circ)$, $1.5 (\vartriangle)$, $4.5 (\blacksquare)$. 
}
\end{center}
\end{figure}

Finally we look at the relative orientation between the polymer end-to-end 
vector ${\bm R}$ and the vorticity vector ${\bm \omega}$. Close to the wall
we find that PDF of the cosine of the angle $\psi$ between ${\bm R}$ and 
${\bm \omega}$ has a peak at zero, see Fig.~(\ref{fig:psi}a).
This implies that the polymers show a weak tendency to lie in the plane perpendicular 
to ${\bm \omega}$. 
However this trend is reversed near the centerline~Fig.~(\ref{fig:psi}b)
where the polymers orient along the vorticity vector.

To summarize the polymers near the wall shows the cleanest trend in their
orientation. They show a strong tendency to line along the stream-wise 
directions.  Weaker trends are seen near the center. The statistics of orientation
of polymers near the center of our flow is very similar to the statistics
of orientation of polymers obtained in homogeneous and isotropic 
flows~\cite{wat+got10}. 
Note however that the orientation effects are
much stronger near the wall than near the centerline.
\section{Conclusions}
\label{sec:conclusion}
We have presented in this paper an extensive numerical study
of the passive Lagrangian polymers in turbulent channel flow.
We have used both linear (Oldroyd-B) and nonlinear (FENE) polymers. 
To understand the statistics of
polymer end-to-end vector it is necessary to know the statistics of the
Finite Time Lyapunov Exponents. For this purpose in addition to 
the polymers we have solved the equation of evolution of infinitismal line
elements in the turbulent flow and calculated the
FTLEs for an inhomogeneous flow. 
We find that the PDF of FTLEs does admit a large deviation expression, 
and we calculate a corresponding Cramer's function. 
Note, however, that the large deviation expression is valid only at very
large times. 
In addition we use the location of the minima of the Cramer's function to 
define our Weissenberg number. Consequently for the FENE 
model we observe coil-stretch transition at $\Wi \approx 1$. 
For the Oldroyd-B model we find that the PDF of polymer
extension shows power-law behavior for $\Wi < 1$. We calculate the exponent
of this power-law using the rank-order method. We also calculate the same
exponent from the Cramer's function using the theory of 
Ref.~\cite{bal+fux+leb00}.
These two different calculations match within error, validating the
theory of Ref~\cite{bal+fux+leb00}.  This shows that the idealizations used
in Ref.~\cite{bal+fux+leb00}, in particular the assumption that in Lagrangian
coordinates the rate-of-strain tensor $\sigma_{\alpha\beta}$ is 
delta-correlated in time is a reasonable approximation at least for
linear polymers below the coil-stretch transition even in the case of 
a realistic flow. For the FENE model we cannot meaningfully calculate the PDF
of polymer extension from the Cramer's function using the results of 
Ref.~\cite{che00} because our numerically calculated Cramer's function is
not accurate enough for this exercise. For the FENE model we find that 
the polymers are more extended near the wall, but the difference 
decreases as Weissenberg number increases far beyond the coil-stretch 
transition. We further find that near the center of the channel 
the orientational statistics of the polymers show similarity 
to orientational statistics obtained for homogeneous 
and isotropic flows~\cite{wat+got10},i.e., they align 
along either of the two largest directions of strain and tend
to orient orthogonal to the third principal direction of strain. 
A much stronger orientational trend is seen near the wall
where the orientations of the polymers are along the 
stream-wise direction.

Although our DNSs involve passive polymers it is possible to 
have insights on polymeric drag reductions from these simulations. 
We can calculate the polymeric stress from our simulations
and add this to the Reynolds stresses to see how they
change the Reynolds-averaged flow equations. It would be interesting
to see how much of drag-reduction can be described by this 
simple approach. Such results will be presented in a future publication. 

\section{Acknowledgements}
\label{acknowledgments}
We thank A.~Brandenburg, O.~Flores, V.~Steinberg, A.~Vulpiani, and D.~Vincenzi for helpful discussions. 
Financial support from the Swedish Research Council under the
grant 2011-5423 and computer time provided by SNIC (Swedish National Infrastructure for 
Computing) are gratefully acknowledged.

\end{document}